\documentclass{vgtc}                          

\ifpdf
  \pdfoutput=1\relax                   
  \usepackage{graphicx}                
  \DeclareGraphicsExtensions{.pdf,.png,.jpg,.jpeg} 
\else
  \usepackage{graphicx}                
  \DeclareGraphicsExtensions{.eps}     
\fi%

\graphicspath{{figures/}{pictures/}{images/}{./}} 

\usepackage{microtype}                 
\PassOptionsToPackage{warn}{textcomp}  
\usepackage{textcomp}                  
\usepackage{mathptmx}                  
\usepackage{times}                     
\usepackage{cite}                      
\usepackage{tabu}                      
\usepackage{booktabs}                  
\usepackage[table,xcdraw]{xcolor}
\usepackage{dblfloatfix}
\usepackage{xcolor}
\usepackage{paralist}

\usepackage{amsmath}
\usepackage{mathptmx}
\usepackage{graphicx}
\usepackage{times}
\usepackage{url}
\usepackage{float}
\usepackage [english]{babel}
\usepackage [autostyle, english = american]{csquotes}
\usepackage{enumitem}
\usepackage{algorithm}  
\usepackage{algorithmic} 
\usepackage{authblk}
\setlist{nolistsep}
\usepackage{booktabs}
\MakeOuterQuote{"}

\usepackage{color}
\newcommand{\squeezeup}{\vspace{-4mm}}

\onlineid{0}

\vgtccategory{Research}

\vgtcinsertpkg

\title{EmbeddingVis: A Visual Analytics Approach to Comparative Network Embedding Inspection}

\author[1]{Quan Li\thanks{e-mail: qliba@connect.ust.hk}}
\author[1]{Kristanto Sean Njotoprawiro\thanks{e-mail: ksean@ust.hk}}
\author[1]{Hammad Haleem\thanks{e-mail: hhaleem@connect.ust.hk}}
\author[2]{Qiaoan Chen\thanks{e-mail: kazechen@tencent.com}}
\author[2]{Chris Yi\thanks{e-mail: chrisyi@tencent.com}}
\author[1]{Xiaojuan Ma\thanks{e-mail: mxj@cse.ust.hk}}
\affil[1]{Department of Computer Science and Engineering, The Hong Kong University of Science and Technology, Hong Kong}
\affil[2]{WeChat, Tencent Technology (Shenzhen) Co., Ltd., Shenzhen, China}

\teaser{
  \squeezeup
  \includegraphics[width=\linewidth]{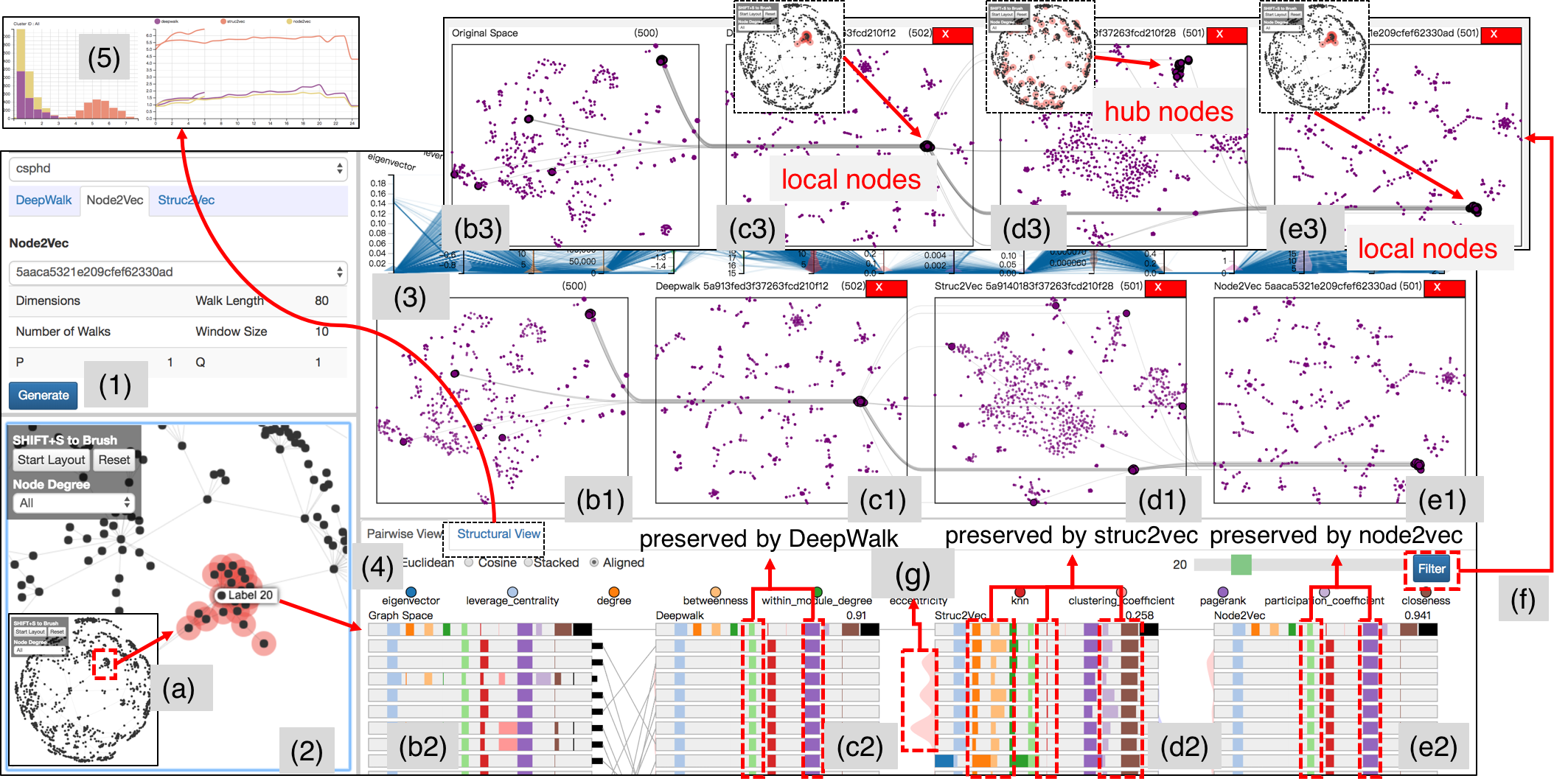}
  \squeezeup
  \squeezeup
  \caption{EmbeddingVis consists of (1) control panel, (2) graph view, (3) cluster transition view, (4) pairwise ranking view and (5) structural view. Using EmbeddingVis to verify preserved metrics at the instance level: (a) users lasso a cluster of nodes and the corresponding nodes are highlighted and linked across different embedding spaces (b1-e1). Clicking one ``hub'' node in this cluster (label 20) in the graph view generates four neighbor ranking lists of this ``hub'' node: (b2) the graph space, (c2) \textit{DeepWalk}, (d2) \textit{struc2vec}, and (e2) \textit{node2vec}. The highlighted rectangles (c2, d2, e2) indicate the preserved metrics by each embedding. (f) Users can click ``Filter'' to highlight the nodes of each ranking list in the cluster transition view (b3-e3). (g) The Euclidean distance between adjacent nodes in \textit{struc2vec} is more volatile compared with that in the other two spaces.}
  \label{fig:teaser}
}

\abstract{Constructing latent vector representation for nodes in a network through embedding models has shown its practicality in many graph analysis applications, such as node classification, clustering, and link prediction. However, despite the high efficiency and accuracy of learning an embedding model, people have little clue of what information about the original network is preserved in the embedding vectors. The abstractness of low-dimensional vector representation, stochastic nature of the construction process, and non-transparent hyper-parameters all obscure understanding of network embedding results. Visualization techniques have been introduced to facilitate embedding vector inspection, usually by projecting the embedding space to a two-dimensional display. Although the existing visualization methods allow simple examination of the structure of embedding space, they cannot support in-depth exploration of the embedding vectors. In this paper, we design an exploratory visual analytics system that supports the comparative visual interpretation of embedding vectors at the cluster, instance, and structural levels. To be more specific, it facilitates comparison of what and how node metrics are preserved across different embedding models and investigation of relationships between node metrics and selected embedding vectors. Several case studies confirm the efficacy of our system. Experts' feedback suggests that our approach indeed helps them better embrace the understanding of network embedding models.%
} 

\CCScatlist{
  \CCScatTwelve{Human-centered computing}{Visualization}{Visualization application domains}{Visual analytics};
  \CCScatTwelve{Human-centered computing}{Visualization}{Visualization design and evaluation methods}{}
}

\begin{document}


\firstsection{Introduction}
\maketitle
\noindent
Data mining and machine learning (ML) enable the discovery of insights into information networks, such as social and paper citation networks. Identifying an appropriate graph representation is necessary prior to conducting any analysis. A direct and conventional representation is the adjacency matrix, whereas edges between nodes are indicated as entries in a matrix. However, this representation frequently suffers due to its quadratic size and potential sparsity and is thus not well suited for many analysis tasks~\cite{salehi2017properties}. Researchers and practitioners have recently found network embedding as a promising and powerful alternative to large network representation~\cite{cai2017comprehensive}. 


\par Network embedding represents a graph in a low-dimensional vector space while preserving as much graph information as possible~\cite{cai2017comprehensive}. Thus, mining tasks performed on the original network, such as node classification~\cite{perozzi2014deepwalk,tang2015line,grover2016node2vec,wang2016structural}, clustering~\cite{li2017attributed}, and link prediction~\cite{liao2017attributed}, can be instead conducted in its embedding space. ML researchers have proposed numerous network embedding techniques that emphasize different items to be preserved in the vector space, such as characteristics of nodes, edges, substructures, or entire graphs. Cai et al.~\cite{cai2017comprehensive} classified existing network embedding techniques into five categories, namely, \textbf{(1) matrix factorization}-based methods, such as singular value decomposition and spectral decomposition (eigen-decomposition)~\cite{cao2015grarep,ou2016asymmetric,huang2017label}; \textbf{(2) deep learning} (DL)-based methods with or without random walk~\cite{perozzi2014deepwalk,grover2016node2vec,tian2014learning,chang2015heterogeneous,wang2016structural}; \textbf{(3) edge reconstruction}-based methods, including maximizing edge reconstruct probability or minimizing distance-based loss or margin-based ranking loss~\cite{tang2015line,lin2015learning}; \textbf{(4) graph kernel}-based methods~\cite{yanardag2015deep}; and \textbf{(5) generative model}-based methods that incorporate latent semantics~\cite{le2014probabilistic,xiao2017ssp}. In this study, we mainly focus on DL-based methods, especially those with random walk, due to their effectiveness and robustness in the absence of feature engineering~\cite{cai2017comprehensive}. Furthermore, these methods present only a few barriers to entry for individuals who aim to explore embeddings and are more general than the others because they consider the intrinsic features of a graph while the others explicitly define a few optimization objectives~\cite{cai2017comprehensive}.

\par Although network embeddings have demonstrated promising performance in various mining tasks, the inner structure of an embedding space and the items preserved in it are not transparent to users for several reasons. \textbf{(1) Abstract Representation.} The basis vectors in a typical embedding, which differ from conventional ``high-dimensional'' vectors, have no explicit meaning~\cite{smilkov2016embedding}. In other words, the actual value of an embedding vector is difficult to interpret and cannot be intuitively mapped to certain graph features. Therefore, vectors from different embedding spaces are not directly comparable, thus obscuring the proper evaluation and leveraging of embedding models for users. Understanding abstract vector representations can be compared with the efforts in ``\textit{understanding what specified input maximizes the activation of a particular element in the neural network}''~\cite{gu2017hidden,salehi2017properties,liu2018deeptracker}. \textbf{(2) Inefficient Exploration.} Users must often undergo trial-and-error processes to manually inspect embedding results due to the stochastic nature of the construction procedure; these processes include random walk-based sampling and non-transparent hyper-parameters. Thus, users demand an intuitive interactive mechanism for the effective comparison and understanding of different embedding results, allocation of computational and human resources, and making informed decisions. \textbf{(3) Shallow-Level Analysis.} Visualization tools such as \textit{Embedding Projector}~\cite{smilkov2016embedding} by Google Brain, have been developed to help interpret embeddings adopted by ML models. By projecting a single embedding space on a $2$D/$3$D display, these tools can facilitate the investigation of the global geometry distribution and local neighbors of a selected vector point. Fine-grained analyses, such as studying and comparing capabilities to retain semantic and (or) structural information across different embedding models, are challenging despite the support of these tools for similarity comparisons that are based on attained embedding vectors. The authors of \textit{Embedding Projector} also indicated that ``\textit{it could be useful to visually compare two embeddings when developing multiple models.}'' To achieve this purpose, nontrivial visualization is required to overcome the difficulties in analyzing and comparing different embedding results. Furthermore, empirical studies on the extent to which a visualization system can resolve breakdowns in ML practitioners' understanding of embeddings and bridge their knowledge with the investigation of embedding results are limited.

\par In this study, we propose \textit{EmbeddingVis}, an interactive visual analytics system that helps ML practitioners understand and compare different embedding models. We conduct an observational study of collaboration partner experts' current practices in social network analysis and identify their primary needs and concerns regarding the use of network embedding. We then investigate what and how node metrics are preserved by selected embedding models through a regression analysis. We also propose the use of the ``average distance vector'' for depicting structural characteristics in studying the capability of an embedding model. On the basis of these objectives, we develop a visualization system to support fine-grained analysis at the cluster, instance, and structural levels. Several cases evaluate the efficacy of our system. Our primary contributions are as follows:
\begin{compactitem}
    \item We identify node metric correlations between the graph space and the embedding space and propose the use of the ``average distance vector'' to depict structural characteristics.
    \item We develop suitable interactive visualizations enhanced with new features to support fine-grained analysis of embedding vectors from the cluster, instance, and structural perspectives.
    \item We showcase an experience of working with ML practitioners and several cases to verify the efficacy of our system.
\end{compactitem}

\section{Related Work}
\noindent
Literature that overlaps with those of this work can be categorized into three groups: embedding model evaluation, explanation of embedding vector space, and visual comparison.

\subsection{Evaluation of Embedding Models}
\noindent
Evaluating embedding models usually requires applying the embeddings to ML tasks. In network compression, Wang et al.~\cite{wang2016structural} and Ou et al.~\cite{ou2016asymmetric} reconstructed the original graph from the embedding space and evaluated the reconstruction error. In node classification, the missing label of a node can be inferred from the links in the network using tagged nodes~\cite{perozzi2014deepwalk,tang2015line,grover2016node2vec,wang2016structural}. In clustering, Li et al.~\cite{li2017attributed} evaluated the effectiveness of \textit{DeepWalk}~\cite{perozzi2014deepwalk} and \textit{LINE}~\cite{tang2015line} and found that they performed similarly. Link prediction can predict future interactions or links that may occur in a growing network, such as prediction of possible friendships~\cite{liao2017attributed}. Visualization also helps in analyzing embedding results. For example, the effectiveness of \textit{DeepWalk} is illustrated by visualizing the Zachary's Karate Club network~\cite{perozzi2014deepwalk}. \textit{LINE} visualizes the \textit{DBLP} co-authorship network and shows that \textit{LINE} can cluster authors from the same field~\cite{tang2015line}. However, all these methods simply project the embedding space to a 2D plane. Although \textit{Embedding Projector} helps explore neighborhoods for individual nodes~\cite{smilkov2016embedding}, fine-grained analysis is still limited.

\subsection{Explanation of Embedding Vector Space}
\noindent
Interpreting embedding space has recently attracted researchers' attention. Dimensionality reduction techniques, such as t-distributed stochastic neighbor embeddings (\textit{t-SNE})~\cite{maaten2008visualizing}, principal component analysis (\textit{PCA}) and Multidimensional scaling (\textit{MDS}), are used to create $2$D embedding for exploring overall structures and linear relationships~\cite{nishana2013graph}. For example, Liu et al.~\cite{liu2018visual} introduced a new embedding to visualize semantic and syntactic analogies and thus determine whether the resulting view captures significant structures.
\par Similarly, network embedding represents each graph node as a low-dimensional vector~\cite{perozzi2014deepwalk}. Although the decomposition method based on eigenvalue provides some formal guarantee for the preserved node properties, methods based on random walk are essentially random and heavily dependent on hyper-parameter settings~\cite{salehi2017properties}. The embedding vector and the internal structure of the embedding space must still be explained intuitively. Rizi et al.~\cite{salehi2017properties} cast the explanation of embedding vectors into a problem of learning to rank these vectors using \textit{RankSVM}. However, the authors considered only the ranking labels of nodes and ignored the concrete similarity values, and the experimental accuracy was only 60\%. Gu et al.~\cite{gu2017hidden} proposed a flow structure-based metric to inspect the embedding space behind the random walk algorithm. Their results showed that the implicit metric space constructed by the flow distance is similar to the \textit{DeepWalk} and \textit{node2vec} embedding space. Unlike the above-mentioned authors, we leverage regression analysis to capture what and how node metrics are preserved by embedding models and further visually identify their importance at the instance level. We also propose a novel approach to leveraging embedding vectors for depicting structural characteristics.

\subsection{Comparative Visualization}
\noindent
Visual comparison of embeddings belongs to the topic of comparative visualization, which is a basic and common visualization task. Gleicher et al.~\cite{gleicher2011visual} summarized three categories of comparative visualization, namely, juxtaposition (i.e., side-by-side), superposition, and explicit encoding (i.e., visual display of differences or correlations). Many visualization methods have been proposed. For example, two juxtaposed identical objects are linked through \textit{VisLink}~\cite{collins2007vislink}. Alper et al.~\cite{alper2013weighted} used the node-link graph and the adjacency matrix to compare different connectivity data in the weighted graph form in brain connectivity analysis. \textit{Lineup}~\cite{gratzl2013lineup} supports sorting items on the basis of multiple heterogeneous attributes with various scales and semantics. This method also compares multiple alternative rankings on the same groups of items. In this work, we mainly use the juxtaposition style to allow users to conduct comparative analysis.

\section{Background and Observational Study}
\subsection{About Network Embedding}
\noindent
A sensible embedding should make key information more accessible for intended tasks than extracting that directly from the original data~\cite{cai2017comprehensive}. In particular, the capability to preserve graph homogeneity and (or) structural functionality determines the quality of a network embedding algorithm. Here, we briefly introduce the network embedding algorithms showcased in this paper, namely, \textit{DeepWalk}~\cite{perozzi2014deepwalk}, \textit{node2vec}~\cite{grover2016node2vec}, and \textit{struc2vec}~\cite{ribeiro2017struc2vec}. \textit{DeepWalk} is the first to apply the idea of \textit{word2vec}~\cite{mikolov2013efficient} to network embedding. It uses a random walk to sample node sequences, which correspond to word sequences (sentences) in a document, and embeds the nodes using the \textit{word2vec} learning algorithm. The framework of \textit{node2vec} is similar to that of \textit{DeepWalk}, except that \textit{node2vec} provides a trade-off between breadth-first-search (BFS) and depth-first-search (DFS) strategy in the random walk process. Unlike the main ideas of preceding models, that of \textit{struc2vec} is built on the observation that structural similarity does not necessarily depend on hop counts; neighboring nodes can be different and distant nodes can be similar. Without using node or edge attributes, \textit{struc2vec} employs a hierarchy to encode the structure of nodes and conducts a weighted random walk down a multi-layer graph representation of the hierarchy to generate the structural context of every node.

\subsection{Experts' Conventional Practice and Bottlenecks}
\noindent
To understand how (well) network embedding gets used in practice, we worked with a team of experts from WeChat\footnote{https://www.wechat.com/en/}. They were two ML practitioners (E.1-2) and two data scientists (E.3-4). A considerable part of their job is to analyze and predict social influence~\cite{hsu2008acceptance} and its diffusion process on a social network of interest, and develop applications such as social marketing accordingly. They have identified that both \textbf{pairwise features} (impact of friends on a user) and \textbf{structural features} (the topological structure of these active neighboring friends)~\cite{ugander2012structural} are important for capturing the complicated mechanism of social influence. They try to quantify and incorporate such knowledge into ML models via feature engineering to predict how possible a user would conduct a specified behavior.

\par Conventionally, the experts take a graph-based heuristic approach to encoding \textbf{pairwise features} in their ML models, such as the number of common friends of two users (i.e., more shared friends meaning stronger influence on each other) and the expected length of a random walk starting from one user to the other (i.e., shorter length suggesting a stronger influence down the path). Similarly, they employ heuristic statistics, such as the number of components and communities, to quantify \textbf{structural features}. Although these graph-based measures do capture some useful information in the graph, the computational time grows rapidly as the network size increases. Besides, feature engineering is a task-dependent process, and its outcome of one application may not be generalizable to other tasks. Envisioning an effortless method, the experts attempted to vectorize the factors for social influence using network embedding.

\par However, the experts encountered the following issues when trying to engage network embedding in their ML models. \textbf{(1) Understanding Obstacles.} The experts design several metrics, such as \textit{k-core} and \textit{betweenness}, to represent node features. These metrics are insufficient for depicting entire networks, but they are particularly easy to understand. In contrast, the embedding vectors are rather abstract as dimensions in a typical embedding space do not have a particular meaning~\cite{smilkov2016embedding}. ``\textit{Some embedding models are still like a baffling mystery to laymen,}'' said E.3. ``\textit{If we know what specific graph information is captured by an embedding, then we would be more confident in using it,}'' said E.4. \textbf{(2) Inconvenient Comparison.} There is no intuitive way to compare the quality of various embedding models, largely due to the stochastic nature of their construction process and the non-transparent hyper-parameters involved. Traditionally, the experts conduct trial-and-error to examine different embeddings on some manually assembled input cases, thereby consuming considerable time. Thus, they demand a more intuitive and interactive technique of inspecting embedding than the conventional method. \textbf{(3) Limited Analysis.} The experts commonly project hundreds of dimensions to an approachable $2$D/$3$D space through dimensionality reduction and visually search for any correlation or pattern that ``\textit{well fit with our understanding of how the model should and does work,}'' said E.1. However, unlike \textit{word2vec}, in which the pairwise word similarity can be easily obtained through their literal meanings, no universal measurement is available to define graph node similarity. Furthermore, calculating pairwise similarity on the basis of the attained embedding vectors is easy, whereas fine-grained analysis, such as comparing capabilities to retain semantic and (or) structural information among embedding models, is challenging.

\subsection{Experts' Needs and Expectations of Embedding}
\noindent
To ensure that the ontological structure of our approach fits well into domain tasks, we interviewed the experts (E.1-4) in two separate sessions to identify the experts' primary concerns about the use of network embedding and potential obstacles in their path to efficient obtaining knowledge. At the end of the interviews, the need for a visualization system to ground the team's conversation~\cite{li2018multi,livisual2017} with embedding emerged as a key theme among the feedback collected. Despite differences in individual expectation for such a system, certain requirements were expressed across the board.

\par \textbf{R.1 Identifying node metrics preserved by different embedding models or by versions with different hyper-parameters.} All the experts were interested in knowing what node metrics in the original graph may better describe node similarity in an embedding space. Such information can help them interpret the meaning of ``neighbor'' and ``community'' in the embedding space, and ultimately determine whether conducting operations like clustering on the resulting vector representations of nodes is appropriate for their intended task. 
\par \textbf{R.2 Analyzing the capability of embedding vectors to retain assorted structural characteristics.} E.1 mentioned that ``\textit{network embedding result only makes sense when we conduct a pairwise node comparison,}'' and E.3 commented that ``\textit{a single node's embedding vector itself provides no additional structural information derived from the original topological space.}'' Therefore, they were interested in determining whether leveraging embedding vectors to describe structural characteristics of interest is practical and whether such a capability differs across various embedding models.
\par \textbf{R.3 Exploring cluster geometry of nodes in embedding space.} The embedding projection will inevitably create clusters. Therefore, E.3-4 expressed a desire to observe the global geometry of the embedding space to facilitate cluster findings and understanding.
\par \textbf{R.4 Exploring the neighbors of a particular node.} A conventional method for E.3-4 to assess the output quality of an embedding was confirming whether the nearby neighbors of a node in the vector space were semantically related. Exploring the neighbors of a given node in $2$D and the original embedding space are both important because the $2$D embedding may distort neighborhood information.
\par \textbf{R.5 Analyzing pairwise node similarity.} Considering there is no one single universal definition of node similarity, all the experts required to show the various relatedness of nodes in a pairwise manner. For example, E.2 said that ``\textit{if two nodes show similar features in terms of topological structures or comparable centralities, they may be accounted as similar nodes.}''
\par \textbf{R.6 Highlighting nodes simultaneously in different spaces.} All the experts were interested in linking the points in the embedding spaces with their corresponding nodes in the original graph for intuitive observation and comparison, so that they could observe the nodes of interest and their contexts in different embedding spaces.

\begin{table*}[h]
\squeezeup
\centering
\caption{The average ranking of feature importances of node metrics for different embedding models on all datasets.}
\label{tab:feature_importances}
\resizebox{\textwidth}{!}{%
\begin{tabular}{ccccccccc}
\hline
\textbf{} & {DeepWalk} & {p=0.004, q=0.004} & {p=1, q=0.004} & {p=256, q=0.004} & {p=256, q=256} & {p=1, q=1} & {p=256, q=1} & {struc2vec} \\
\hline
\textit{{Degree}} & 0.33\% & 0.23\% & 0.24\% & 0.30\% & 0.22\% & 0.21\% & 0.27\% & \cellcolor[HTML]{9AFF99}{\color[HTML]{333333} 70.98\%} \\
\textit{{Betweenness}} & 4.30\% & 4.10\% & 4.07\% & 3.89\% & 4.62\% & 4.03\% & 4.01\% & 5.18\% \\
\textit{{Leverage\_Centrality}} & 6.88\% & 6.43\% & 5.14\% & 4.84\% & 6.09\% & 6.55\% & 5.22\% & 3.00\% \\
\textit{{KNN}} & 8.82\% & 7.51\% & 5.79\% & 5.53\% & 8.09\% & 7.86\% & 5.89\% & 1.69\% \\
\textit{{Closeness}} & 9.82\% & 8.35\% & 8.00\% & 7.69\% & {\color[HTML]{333333} 8.20\%} & 7.80\% & 8.16\% & 1.77\% \\
\textit{{PageRank}} & 14.50\% & 12.89\% & 10.53\% & 10.60\% & 13.65\% & 13.15\% & 9.50\% & 11.00\% \\
\textit{{Within\_Module\_Degree}} & \cellcolor[HTML]{9AFF99}{\color[HTML]{333333} 41.93\%} & \cellcolor[HTML]{9AFF99}{\color[HTML]{333333} 50.02\%} & \cellcolor[HTML]{9AFF99}{\color[HTML]{333333} 57.73\%} & \cellcolor[HTML]{9AFF99}{\color[HTML]{333333} 58.82\%} & \cellcolor[HTML]{9AFF99}{\color[HTML]{333333} 49.29\%} & \cellcolor[HTML]{9AFF99}{\color[HTML]{333333} 49.62\%} & \cellcolor[HTML]{9AFF99}{\color[HTML]{333333} 58.79\%} & 2.03\% \\ \hline
\end{tabular}
}
\squeezeup
\end{table*}

\section{EmbeddingVis}
\begin{figure}[h]
\vspace{-2mm}
    \centering
    \includegraphics[width=\linewidth]{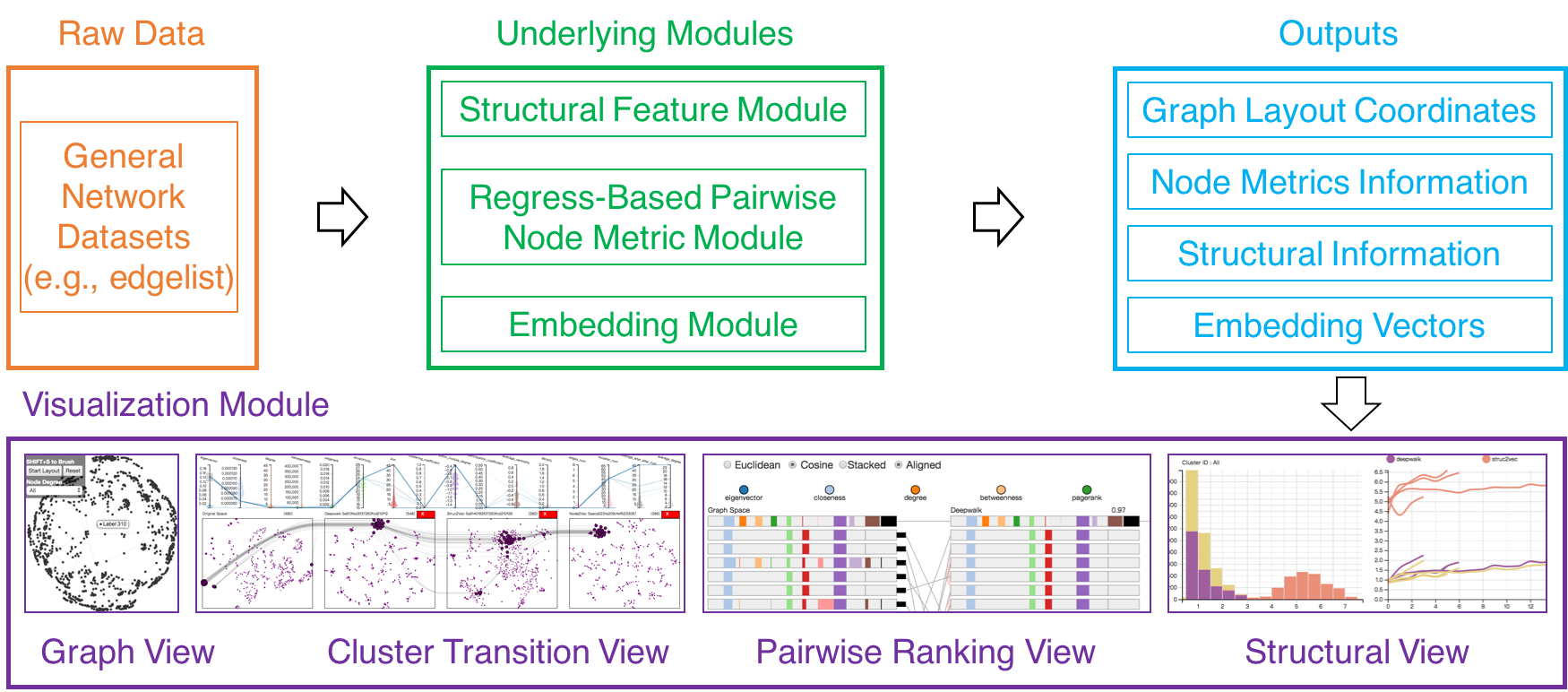} 
    \squeezeup
    \caption{The network dataset is processed by three underlying modules (embedding module, regression-based pairwise node metric module, and structural feature module). The outputs (graph layout coordinates, node metrics information, structural information and embedding vectors of nodes) are then fed into visualization module.}
    \label{fig:pipeline}
\end{figure}

\noindent
\textit{EmbeddingVis} is a web-based visual analytics system that assists users in studying and comparing embedding spaces. This system was developed under the full-stack \textit{MEAN.js} (i.e., \textit{MongoDB}, \textit{ExpressJS}, \textit{AngularJS}, and \textit{Node.js}) framework. As shown in Fig.~\ref{fig:pipeline}, to start the exploration, users can select a network dataset and three underlying modules, namely, the \textit{embedding module}, the \textit{regression-based pairwise node metric module}, and the \textit{structural feature module}, further process this dataset. To be specific, the \textit{embedding module} applies the network embedding models to the dataset. The \textit{regression-based pairwise node metric module} extracts the node metrics from the data and generates the feature importance for each metric. The \textit{structural feature module} extracts the `focal node'-`neighborhood' information of the network. The outputs of these modules are fed into the \textit{visualization module}, which provides three levels of analysis, namely, the cluster level as the \textit{cluster transition view}, the instance level as the \textit{pairwise ranking view}, and the structural level as the \textit{structural view}. This module also provides rich real-time interactions that enable the experts to effectively inspect the embedding results in a fine-grained comparative manner.

\subsection{Regression-based Pairwise Node Metric Analysis}
\noindent
We fit the graph space to the embedding space using regression analysis (R.1) to understand what and how node metrics are potentially preserved by embedding vectors. Regression analysis has been long used to model the relationship between variables and estimate how a dependent variable responds to changes~\cite{armstrong2011illusions}. We regard each designated metric as an observed variable and use its underlying combination to regress the embedding space. We then weigh each metric in terms of its contribution to the prediction by using the feature importance. We approximate the similarity defined by pairwise embedding vectors through a regression model based on a list of selected metrics as 
$
sim_{(Y_u, Y_v)} \sim reg(w_1 \cdot p_1(u, v), w_2 \cdot p_2(u, v), ..., w_k \cdot p_k(u, v))
$, where $sim$ computes the pairwise similarity of $u$ and $v$ based on their embedding vectors $Y_u$ and $Y_v$, and $p_i$ computes the pairwise similarity of $u$ and $v$ based on the studied metric $i$.

\par \textbf{Node metrics.} Numerous measures have been proposed to highlight a node's situation in its network. A few of these measures are on the local level, while others are on the global level~\cite{labrini2015graph}. Apart from the well-known measures (1) \textit{Degree}, (2) \textit{Eccentricity}, (3) \textit{Closeness}, (4) \textit{Betweenness}, (5) \textit{Eigenvector}, (6) \textit{PageRank}, and (7) \textit{Clustering Coefficient}, the following typical state-of-the-art measures are adopted in this study: (8) \textit{Average Nearest Neighbors Degree (knn)}: $\frac{1}{k_i}\sum_ja_{ij}k_j$ studies the surroundings where a node is situated. This measure is based on the observation that a node with low degree may be surrounded by nodes of various sizes, which may directly influence its own centrality and growth potential~\cite{labrini2015graph}. (9) \textit{Within Module Degree}: $({K_i - K_{S_i}})/{\sigma_{K_{S_i}}}$, where $K_i$ is the degree of node $i$ in the community $S_i$, $K_{S_i}$ is the average degree of all nodes in the community $S_i$, and $\sigma_{K_{S_i}}$ is the standard deviation, indicates how well connected a node is to other nodes in the same community. (10) \textit{Participation Coefficient}: $1-\sum_{m=1}^M(\frac{k_{i,m}}{k_i})^2$, where the term $k_{i,m}$ denotes the number of connections between node $i$ and other nodes within community $m$, and $k_{i,m}/k_i$ indicates the ratio of connections a node has within its own community. (11) \textit{Leverage Centrality}: $\frac{1}{k_i}\sum_{N_i}\frac{k_i - k_j}{k_i + k_j}$ considers the degree of a node ``relative to its neighbors''. A feature signature vector $V1$ for each node is then constructed using these metrics and $V1$ will be leveraged in Section 4.3.1.

\par \textbf{Identifying preserved node metrics.} We conduct an experiment to approximate the similarity of two embeddings $Y_u$ and $Y_v$ by the preceding node metrics. We compute the correlation between pairwise nodes in the embedding and metric spaces primarily to identify what and how node metrics are preserved by different embedding models. We attain the preceding defined node metrics of each node in a given network. For each pair of nodes, we compute the Euclidean distance in the metric space. We also obtain the pairwise Euclidean distance by the embedding vectors generated from the embedding models (\textit{DeepWalk}, \textit{node2vec}, and \textit{struc2vec} in this paper) to understand which nodes remain close to each other after embedding.

\par \textbf{Dataset.} We evaluate our regression approach on several real-world datasets, namely, (1) \textit{csphd}~\cite{parberry1995sigact}, which contains the ties between Ph.D. students and their advisors in theoretical computer science and has $1025$ nodes and $1043$ edges; (2) \textit{citeseer}~\cite{lawrence1999digital}, which contains $3312$ publications and $4732$ connections among them; (3) \textit{wiki}~\cite{sen2008collective}, which contains $2405$ web pages with $17981$ links and is denser than \textit{citeseer}; (4) \textit{email}~\cite{yin2017local}, a dataset from a large European research institution with $1005$ nodes and $25571$ edges.

\par \textbf{Parameter settings.} Following the same parameters used for comparing different embeddings~\cite{perozzi2014deepwalk, grover2016node2vec, ribeiro2017struc2vec, salehi2017properties}, we set the parameters for all models as follows: window size $10$, embedding dimension $128$, number of random walks for each node $10$, and walk length $80$. For \textit{node2vec}, two hyper-parameters $p$ and $q$ control the sampling strategy of random walks. We consider three parameter permutations of $.004$, $1$, and $256$ in our experiment. Therefore, nine groups of parameters are available. The parameters in \textit{struc2vec} are the same as those in \textit{DeepWalk}.

\par
\begin{figure}[h]
\squeezeup
    \centering
    \includegraphics[width=\linewidth]{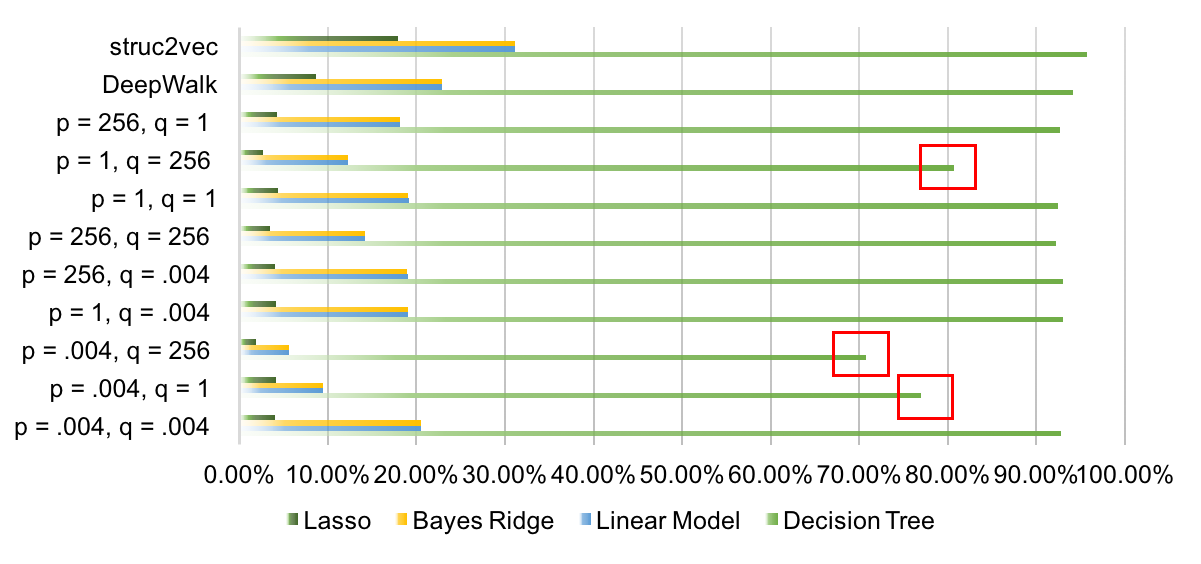} 
    \squeezeup
    \squeezeup
    \caption{X axis is the average $R^2$ score over all datasets. Y axis indicates different embeddings. Decision Tree overwhelms the others and three versions of \textit{nodevec} have a comparatively lower performance.}
    \label{fig:avgper}
\end{figure}

 \textbf{Experiments.} We compare the performances of four regression models, namely, \textit{Decision Tree Regression}, \textit{LinearRegression}, \textit{BayesRidge}, and \textit{Lasso}, to understand the underlying governing equation that helps associate an individual metric with the network embedding techniques. We split the data into 80\% training subset and 20\% test subset and apply these regression models on the previously mentioned datasets. Fig.~\ref{fig:avgper} shows the average $R^2$ scores\footnote{https://en.wikipedia.org/wiki/Coefficient\_of\_determination} (\textit{the coefficient of determination statistically measuring how well the regression line approximates the real data}) of these regression models on all datasets, and the results indicate that \textit{Decision Tree Regression} can predict the pairwise node distance in the embedding space with sufficiently high accuracy. We then leverage \textit{Decision Tree Regression} and extract the feature importance for each node metric. We multiply the feature importance with the $R^2$ score of each embedding model and rank each of these features. Only the embedding models with $R^2$ scores higher than $0.70$ are selected. Therefore, a prominent feature extracted from an embedding model with a high $R^2$ score represents a high correlation between the distance of two nodes in the embedding space and the selected node metric. After ranking all features, we select all different metrics with scores higher than the $50^{th}$ percentile. These metrics are presented in Table~\ref{tab:feature_importances}. Evidently, \textit{within module degree} overwhelms the other metrics in \textit{DeepWalk} and \textit{node2vec}, while \textit{struc2vec} mostly preserves \textit{degree}.

\begin{figure*}[h]
    \centering
    \squeezeup
    \includegraphics[width=\linewidth]{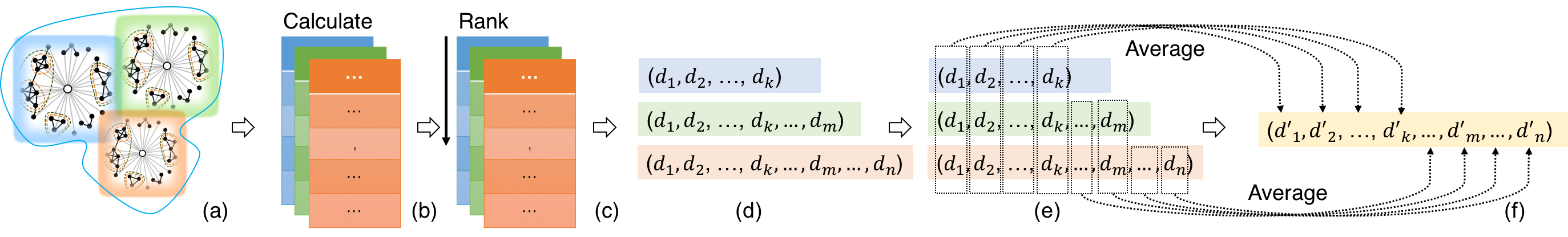} 
    \squeezeup
    \squeezeup
    \caption{Structural feature measurement. (a) Suppose we have three ``focal node''-``neighborhood'' networks in a cluster. For each network, (b) we obtain pairwise links between the ``focal node'' and its neighbors, (c) calculate and rank the Euclidean distance based on the embedding vectors of the links' nodes. (d) We use the distance vectors to represent the structural information of each ``focal node''-``neighborhood'' network. (e) We average the value on each dimension, and (f) is the ``\textbf{average distance vector}'' that represents the structural information of this cluster.}
    \label{fig:structures}
    \squeezeup
\end{figure*}

\begin{figure*}[!b]
    \centering
    \squeezeup
    \includegraphics[width=\linewidth]{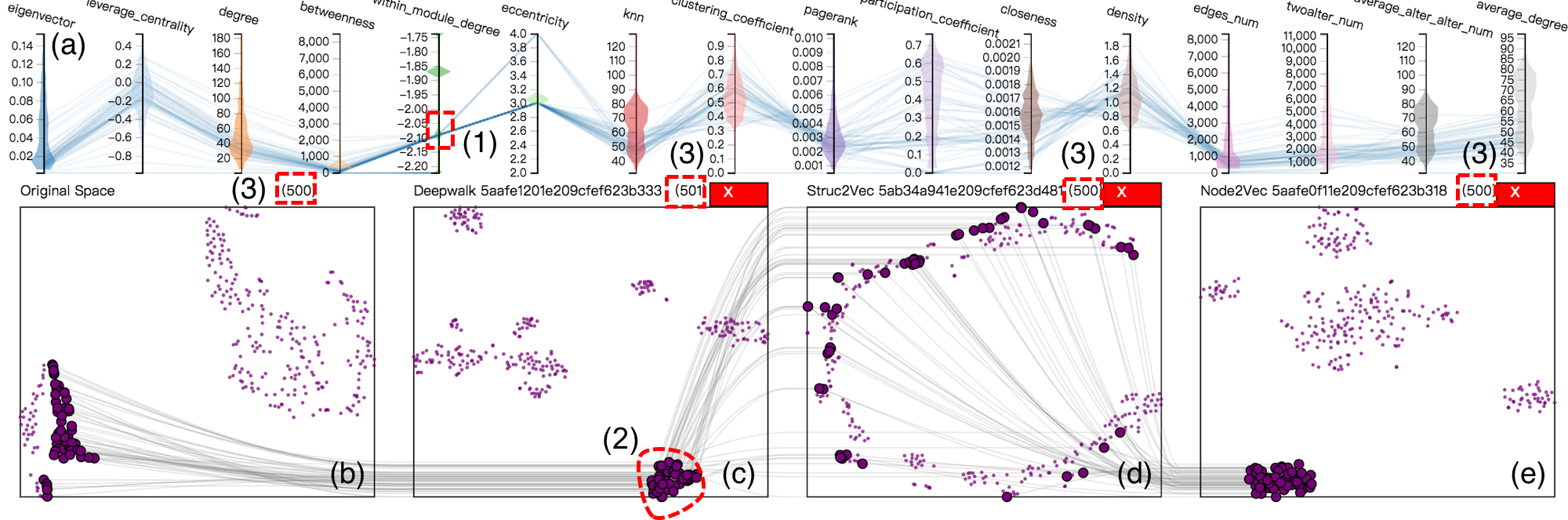} 
    \squeezeup
    \squeezeup
    \caption{The cluster transition view consists of (a) node metric parallel coordinates plot and \textit{t-SNE} projection spaces by (b) the original graph, (c) \textit{DeepWalk}, (d) \textit{struc2vec}, and (e) \textit{node2vec}. Users can filter nodes by either (1) brushing a certain range of values in an axis or (2) lassoing nodes in any \textit{t-SNE} space, and all the identical nodes will be connected by curves. (3) The number indicates the iteration number of \textit{t-SNE}. When clicking the red box with $X$ mark, the corresponding embedding space will be removed.}
    \label{fig:cluster_transition}
\end{figure*}

\subsection{Structural Feature Measurement}
\noindent
In this section, we study and compare the capability of embedding models to depict the structural characteristics (R.2). The motivation is that the embedding results can be understood only when we compare one embedding vector with another, and a single embedding vector alone does not provide any structural information of the graph. Unlike the previous pairwise node similarity analysis, here we consider the structure of the `focal node'-'neighborhood'.

\par Our approach comprises two steps. First, on the basis of studies~\cite{wu2016egoslider,li2017visual}, we select the following metrics that can represent the `focal node'-`neighborhood' characteristics, namely, (1) \textit{number of alters of the focal node} (\textit{degree}): $|V_u - {u}|$; (2) \textit{number of edges among neighbors} (\textit{edges\_num}): $|E_u| - n_u$; (3) \textit{density of the neighborhood} (\textit{density}): ${L_u}/{(n_u(n_u-1)/2)}$; (4) \textit{number of focal node's $2$-step neighborhood} (\textit{twoalter\_num}): $|{w \in V_v, v \in V_u, w \notin V_u}|$; (5) \textit{average number of neighbors of neighbors' networks} (\textit{average\_alter\_alter\_num}): $avg(n_v), v \in V_u^t - {u}$; (6) \textit{average degree of neighborhood} (\textit{average\_degree}): ${\sum_{i=1}^{1+n}n_i}/{1+n_u}$; (7) \textit{clustering coefficient}: ${|\{e_{jk}:v_j,v_k \in N_i, e_{jk} \in E\}|}/{k_i(k_i-1)}$. A feature signature vector $V2$ of a focal node is then constructed using these seven metrics. Second, we obtain the links from the focal node to its neighbors and calculate the Euclidean distances on the basis of the embedding vectors of the links' nodes. The rationale is that a graph's structure is decided by node position, which can be described by the distribution of distances between the links' nodes~\cite{ugander2012structural}.

\par We cluster the `focal node'-`neighborhood' structures by kmeans on the basis of the preceding metrics mentioned. We compute the distance between the `focal node'-`neighborhood' structures using Canberra Distance~\cite{berlingerio2012netsimile}: $dCan(U, V) = \sum_i^n(|U_i - V_i|/(U_i + V_i))$, where $U$ and $V$ represent the feature signatures of two `focal node'-`neighborhood' structures. We replace the distance function in $k$-means with Canberra Distance because its computation is inexpensive, it is sensitive to slight changes, and it normalizes the absolute difference of individual comparisons~\cite{wu2016egoslider}. As shown in Fig.~\ref{fig:structures}, we then obtain the `focal node'-`neighborhood' structures for each cluster and attain the pairwise links between the focal node and its neighbors for each of them. Next, we sort the distances on the basis of the embedding vectors of the link nodes to create a \textbf{distance vector} for each `focal node'-`neighborhood' structure. Following this pipeline, we average the values of each dimension of the distance vectors in this cluster and obtain the ``\textbf{average distance vector}'' that represents this cluster. This method is adopted to determine how the values of each dimension of the average distance vector change. The use of the feature signature vector $V2$ and the corresponding visual designs are introduced in Section 4.3.3.

\subsection{Visualization}
\noindent
The basic design principle behind \textit{EmbeddingVis} is leveraging or augmenting familiar visual metaphors to enable experts to focus on analysis~\cite{mclachlan2008liverac}. We strictly follow the mantra ``\textit{overview first, zoom and filter, then details-on-demand,}''~\cite{shneiderman2003eyes} which guides users in clearly exploring the embedding vectors. On the basis of these principles and the preceding requirements mentioned, we develop three visualizations that allow the outputs of embedding models to be easily inspected at the cluster, instance, and structural levels. Specifically, we design a cluster transition view to illustrate the neighboring changes in different embedding results, a lineup-based pairwise ranking view to facilitate the exploration of node neighbors and a structural view to show the distance distribution of `focal node'-`neighborhood' structures across different embedding models.

\subsubsection{Cluster-Level as Cluster Transition View}
\noindent
Dimensionality reduction techniques, such as \textit{t-SNE}, \textit{PCA} and \textit{MDS}, which create low-dimensional embeddings and preserve local similarities to convey neighborhood structure~\cite{heimerl2018embeddings}, have been widely adopted to explore or illustrate the patterns inside a model's learned representation features and generate an overview of embedding results~\cite{rauber2017visualizing}. For example, Liu et al.~\cite{liu2018visual} adopted projection to demonstrate the model's effectiveness in preserving word-level semantic relationships such as analogy and cluster gatherings. Although it is popular to view local neighborhoods in embeddings, projecting in the last step alone provides only an overall structure of the learned representations. Analyzing and comparing several embeddings at the cluster level still requires a more elaborated overview.

\par \textbf{Visual Encoding and Interaction.} We design a cluster transition view (Fig.~\ref{fig:cluster_transition}) to analyze the distribution of a collection of nodes in different embedding spaces (R.1, R.3). This view contains a parallel coordinate plot (\textit{PCP}) and a \textit{t-SNE} embedded transition diagram (Fig.~\ref{fig:cluster_transition}(a, b-e)). The \textit{PCP}~\cite{inselberg1987parallel} displays the metric value distribution of the entire network nodes and the histogram distribution of each metric on each axis. Users can brush on axes to filter nodes. The \textit{t-SNE} embedded transition diagram connects the \textit{t-SNE} projections from different spaces, that is, the original graph space as the first, and three selected embedding spaces as the second, third, and fourth (In this paper, we set the number of comparative embedding spaces to three). We construct the original graph space $2$D embedding using \textit{t-SNE} projection on the basis of the feature vector $V1$ (see node metrics in Section 4.1) of each node. The three embedding spaces can be dynamically removed and added (Fig.~\ref{fig:cluster_transition}(red box with X mark)), thus users would not be overwhelmed by different combinations or orders of the embedding spaces. Users can also lasso nodes on any \textit{t-SNE} projection space and all identical ones will be connected via curves. Furthermore, the \textit{PCP} will highlight the lassoed nodes to show their metric value distributions.

\par We select \textit{t-SNE} as the dimensionality reduction technique because it shows superiority in generating 2D projection that ``\textit{can reveal meaningful insights about data, e.g., clusters and outliers}''~\cite{kim2016pixelsne}. It is more visually interpretable results than naive eigen-analysis, and depending on the distribution, more intuitive than \textit{MDS} results, which preserve global structure more at the expense of local structure retained by \textit{t-SNE}~\cite{heimerl2018embeddings}. Regarding the design, initially, we use interactive animation to track the transition of nodes between two embedding spaces. However, our collaboration experts reported that tracking the nodes in an animated manner requires a mental map comparison, which is ``\textit{demanding, especially when involving simultaneous tracking of additional nodes.}'' Therefore, we develop the cluster transition view based on the juxtaposition comparative visualization.

\begin{figure*}[h]
    \centering
    \squeezeup
    \includegraphics[width=\linewidth]{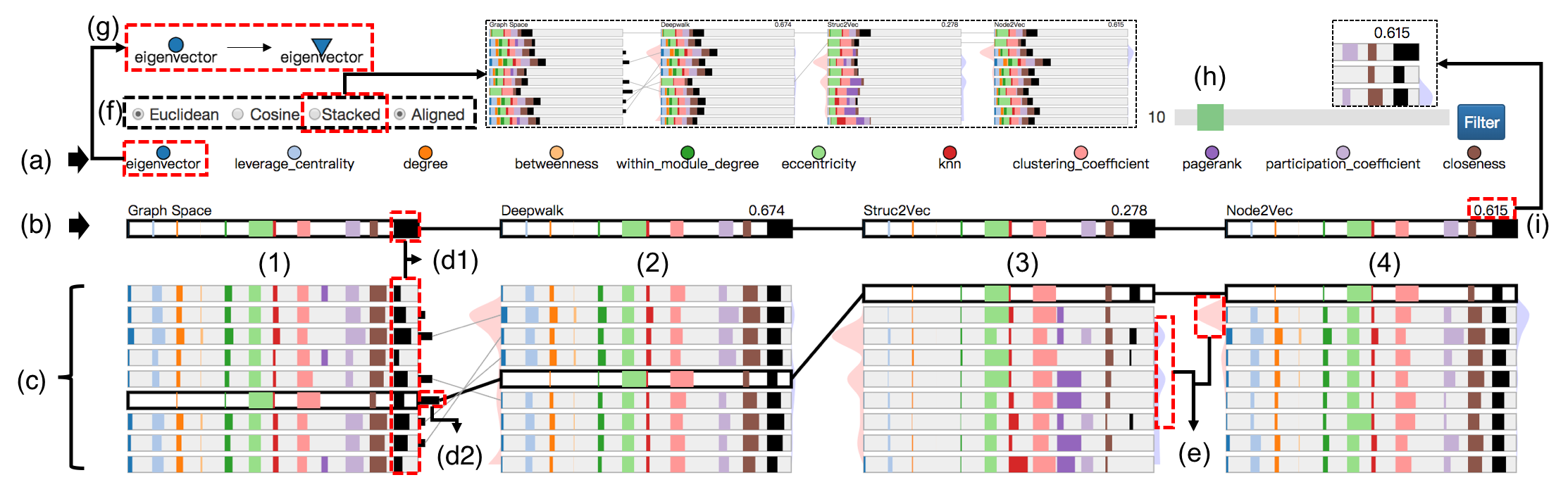} 
    \squeezeup
    \squeezeup
    \caption{Pairwise ranking view contains four ranking lists in (1) the original graph and (2, 3, 4) three selected embedding spaces. (a) 11 metrics are displayed as legends. (b) The top row in each ranking represents the selected node. (c) Top N neighbors of the selected node. (d1) Number of shared friends between the neighbor and the selected node. (d2) The length of the black bars indicate how many embedding spaces contains the corresponding node. (e) Curve areas indicate the similarity changes of two adjacent nodes. (f) Control options for distance calculation and layout. (g) Users can click a metric to order the graph nodes based on the clicked metric. (h) Users can control the length of displayed neighboring ranking list. (i) The NDCG score measures the ranking quality of each ranking list.}
    \label{fig:pairwise_ranking}
    \squeezeup
\end{figure*}

\subsubsection{Instance-Level as Pairwise Ranking View}
\noindent
Although 2D displays convey additional information about neighborhood structure, it remains unclear how much of it is noise that is captured in the process as opposed to relevant semantic structure. This makes interpretation, and in particular comparison across embeddings challenging~\cite{heimerl2018embeddings}. Moreover, ML experts often need to explore the properties of an embedding at the instance level. For example, an engineer working on a recommendation system who creates an embedding algorithm for songs might verify that the nearest neighbors of ``\textit{Stairway to Heaven}'' contain ``\textit{Whole Lotta Love}'' rather than ``\textit{Let It Go}''~\cite{smilkov2016embedding}. Applying embedding to rank the neighboring nodes and measuring its performance are also meaningful in network analysis. First, sparsity problems frequently occur in social networks. For example, Bob has $140$ friends, and only $20$ of them frequently communicate with him. For other friends, relying only on scarce communication behaviors to rank intimacy is difficult. Second, in some scenarios, such as communication networks, some persons, such as insurance agents, communicate frequently with Bob only for a short while for business reasons. If we rank the intimacy between Bob and his friends on the basis of their communication frequency, then these ``business acquaintances'' may be assigned unreasonably high rankings. Third, no universal measurement is available for defining node similarity. Thus, exploring the neighboring information can help experts assess the output quality of an embedding because neighbors may have similar features (R.4-5).
\par After discussing with the experts, we adopt the preceding metrics (Section 4.1) to evaluate the neighboring information at the instance level. Users can then conduct a detailed comparison between two neighboring nodes in terms of metrics. However, manually selecting nodes for comparison is cumbersome, especially when the graph is complicated and no clear cues are available for users to decide which nodes to compare. Therefore, we convert pairwise comparison to ranking comparison to quantify different embedding ranking qualities and understand how (well) node metrics are preserved.

\par \textbf{Ranking Measurement.} To measure the ranking quality of different embeddings, we use \textit{Discounted Cumulative Gain (DCG)} which is widely used to measure the efficacy of search engines~\cite{wang2013theoretical}. The \textit{DCG} accumulated at a particular rank position $k$ is defined as 
$
DCG_k = \sum_{i=1}^k\frac{rel_i}{log_2(i+1)} = rel_1 + \sum_{i=2}^k\frac{rel_i}{log_2(i+1)}
$. By standardizing the queries, all relevant documents are sorted in the corpus according to their relative correlation, and the possible \textit{DCG} is generated by the position $k$, which is called \textit{Ideal DCG (IDCG)}. Therefore, we use \textit{normalized discounted cumulative gain (NDCG)}:
$
NDCG_k = DCG_k/IDCG_k
$, where \textit{IDCG} is defined as 
$
IDCG_k = \sum_{i=1}^{|REL|}\frac{2^{rel_i}-1}{log_2(i+1)}
$ and $|REL|$ represents the list of relevant documents (ordered by their relevance) in the corpus up to position $k$.

\par \textbf{Design Criteria.} To facilitate the exploration and comparison of neighboring information of a selected node across different embeddings, the pairwise ranking view should meet the following design criteria: (1) \textit{Encode Ranking Similarity.} Users should be able to rapidly grasp the rankings that are determined by pairwise node similarities. (2) \textit{Encode Ranking Causes.} Users must be able to evaluate the details of each node, such as node metric distribution and similarity ranking, to understand how the rankings are determined. (3) \textit{Compare Rankings Between Models.} Confirming the nearby neighbors of a node is necessary for evaluating and trusting an embedding model. That is, multiple rankings from different embedding models must be placed into context with one another, thereby allowing users to compare multiple models simultaneously. (4) \textit{Interactive Operations.} Interactive mechanisms on the ranking in the original graph space, such as sorting by a certain node metric, should be provided to enable users to determine the difference between embedding models reflected on rankings. Thus, users can witness the response of embedding spaces and the overlaps of neighbors between the original graph space and the embedding space.

\par \textbf{Visual Encoding and Interaction.} Inspired by \textit{Lineup}~\cite{gratzl2013lineup}, we design a pairwise ranking view to help users conduct instance level analysis. We present each node metric (Fig.~\ref{fig:pairwise_ranking}(a)) as a separate column. These columns adopt bars with different colors to represent the normalized values for each node metric. The top row in each ranking always represents the currently selected node (Fig.~\ref{fig:pairwise_ranking}(b)), and the rows below represent the neighbors of the selected node (Fig.~\ref{fig:pairwise_ranking}(c)). We add an additional bar (Fig.~\ref{fig:pairwise_ranking}(d1)) to represent the number of shared friends between the neighboring node and the selected node. To compare the neighboring rankings, we line up the four rankings horizontally (Fig.~\ref{fig:pairwise_ranking}(1-4)), each having its own neighboring node order, and connect identical nodes across the neighboring rankings with lines (linked and highlighted black rectangles in Fig.~\ref{fig:pairwise_ranking}). These rankings are that in the original graph space (\textit{The ranking is based on the neighbors of the selected node in the original graph}) and those from the three embedding spaces (\textit{The rankings are based on the Cosine or Euclidean similarity between the neighbor node and the selected node and determined using the embedding vector from the corresponding embedding model}). We design a small bar (Fig.~\ref{fig:pairwise_ranking}(d2)) and use its length to encode the number of embedding spaces which capture the corresponding node in their neighboring ranking list (Since we support three embedding spaces for simultaneous comparison and if all spaces capture a node in their ranking lists, the length of the bar is $3$). The similarity measures based on Cosine or Euclidean distance are encoded by two curve areas (Fig.~\ref{fig:pairwise_ranking}(e)) that lie on both sides of the columns. Here, we visualize the similarity change of two adjacent nodes, i.e., a peak indicates a considerable change of similarity between the two adjacent nodes. This method can resolve the ranking ambiguity of two nodes with the same similarity. As shown in Fig.~\ref{fig:pairwise_ranking}(f), changes in the option of Cosine or Euclidean distance will influence only the ranking of neighboring nodes in the embedding spaces. Two alignment strategies are provided, namely, the use of stacked and aligned bars. The two strategies can be toggled dynamically. Operations such as sorting by \textit{eigenvector} (Fig.~\ref{fig:pairwise_ranking}(g)) between the selected node and all the other nodes in the whole graph will influence the order in the original graph space, i.e., the first column. We add a filter slider (Fig.~\ref{fig:pairwise_ranking}(h)) that controls the length of neighboring rankings due to limited vertical space. The \textit{NDCG} scores (Fig.~\ref{fig:pairwise_ranking}(i)) on top of each embedding column indicate the ranking qualities of different embeddings.

\subsubsection{Structure-Level as Structural View}
\noindent
To understand the capability of embedding models to depict the structural characteristics, we obtain the links of each `focal node'-`neighborhood' structure from the focal node to its neighbors and calculate the Euclidean distance on the basis of their embedding representations, namely, the ``distance vector''. We then obtain the ``average distance vector'' for each cluster of `focal node'-`neighborhood' structures (Section 4.2). We design a structural view to help visualize the distance distribution of `focal node'-`neighborhood' structure clusters across different embeddings (R.2).

\par
\begin{figure}[h]
    \centering
    \includegraphics[width=\linewidth]{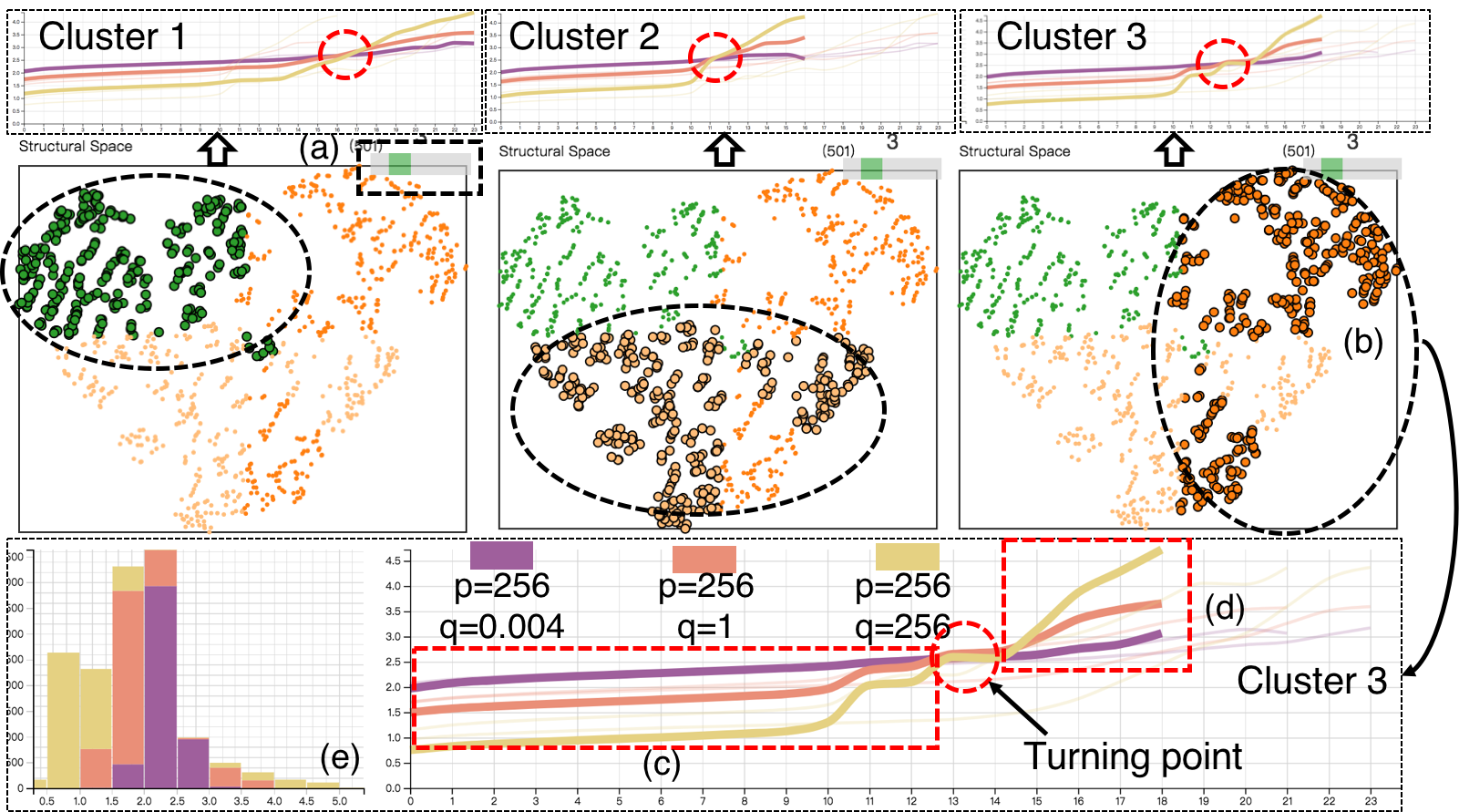} 
    \squeezeup
        \vspace{-3mm}
    \caption{(a) Nodes are clustered into three. (b) Hover on one cluster and observe its ``average distance vector'' curves generated by three versions of \textit{node2vec}. (c, d) A ``turning point'' occurs at the curve of (p=$256$,q=$256$) (indicated by the yellow curve).}
    \label{fig:structure}
\end{figure}

 \textbf{Visual Encoding and Interaction.} As shown in Fig.~\ref{fig:structure}(a), we construct the graph space using \textit{t-SNE} projection of the feature vector $V2$ (Section 4.2) of each `focal node'-`neighborhood' structure. Users can change cluster number in kmeans. Colors in the 2D spaces represent clusters. The stacked bars in Fig.~\ref{fig:structure}(e) show the distribution of Euclidean distance on the basis of embedding vectors from each embedding indicated by a color. X axis shows the distance bins, and the bar length shows the number of distances under each bin. In Fig.~\ref{fig:structure}(c-d), the highlighted curves represent the ``average distance vector'' of a cluster (e.g., cluster 3 in Fig.~\ref{fig:structure}(b)). X axis represents the dimension index of the ``average distance vector'', and Y axis is the value at the corresponding dimension. In other words, the maximum value of X axis indicates the maximum degree of the network.

\subsection{Interactions Among the Views}
\noindent
Apart from the most defining capabilities of \textit{EmbeddingVis}, rich interactions are integrated to catalyze an efficient in-depth analysis (R.6). (1) \textit{Filtering and Highlighting.} Users can select/lasso nodes in the graph view or in the cluster transition view to inspect nodes or areas of interest, and the system will automatically highlight the corresponding information in all the other views. (2) \textit{Linking.} Views are automatically linked through nodal correspondence. The coordinated interactions among these nodes facilitate examination of a variety of information at different granularities, such as the cluster and instance levels, thus forming convenient hypothesis generation and verification. (3) \textit{Animation.} The projection in the cluster transition view and operations in the pairwise ranking view provide an intuitive process of cluster formation and ranking changes.

\section{Evaluation}
\noindent
In this section, we demonstrate system efficacy by conducting two case studies. The first case evaluates node metrics preserved by embedding models. The second case compares the effects of hyper-parameters in different versions of an embedding. Before introducing the two cases, we first report how the experts familiarized themselves with the system by using one of their familiar networks.

\par
\begin{figure}[h]
\centering
\includegraphics[width=\linewidth]{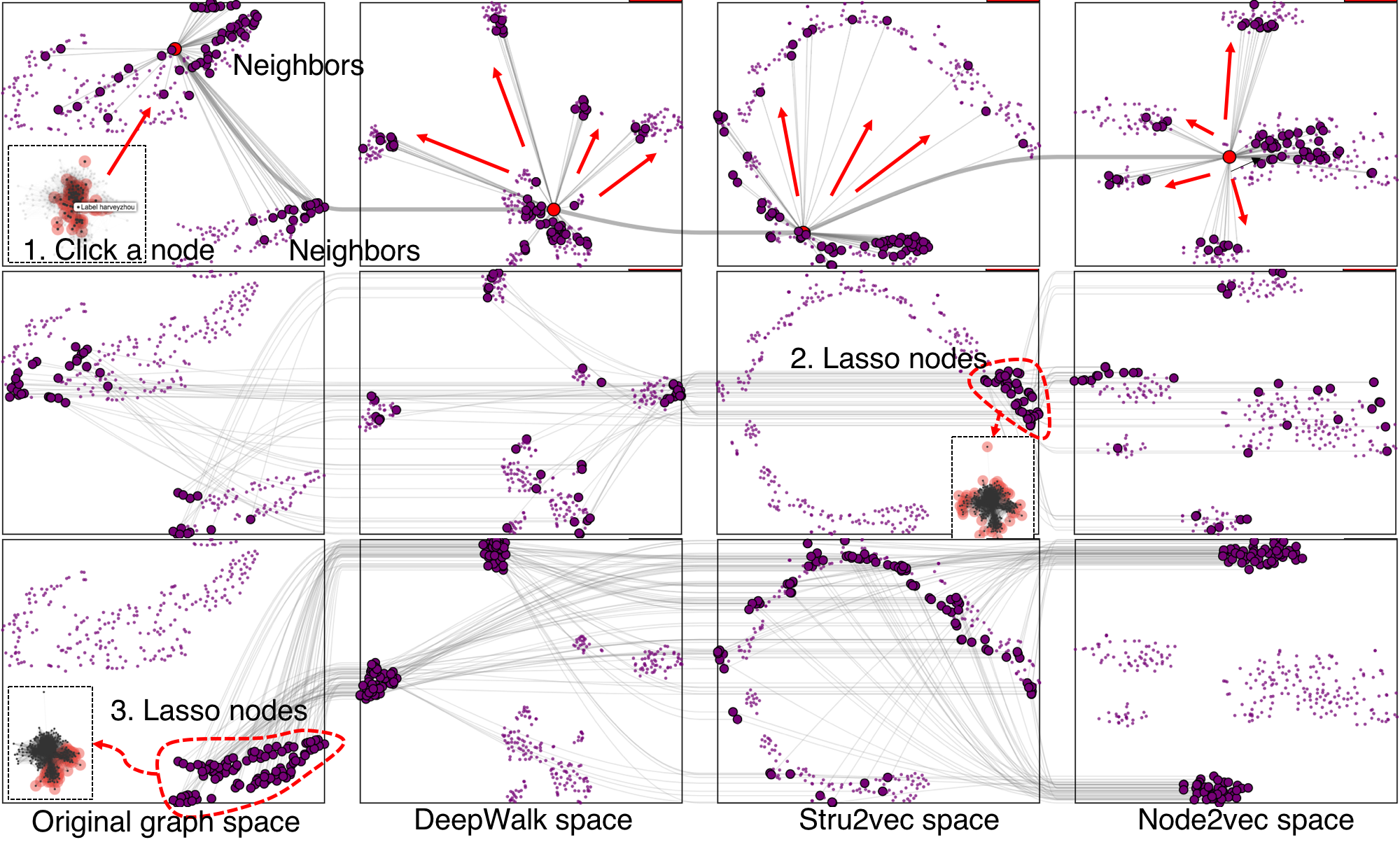} 
\squeezeup
\squeezeup
\caption{(1) Click a node in graph view and the identical ones are linked via a curve (inter-embedding link). Links from this node to its neighbors are also shown (inner-graph links). The widths of the two types of links are different. (2) Lasso several nodes in \textit{struc2vec} and the graph view shows that they correspond to department interns. (3) The lassoed nodes represent two groups of employees.}
\label{fig:case0}
\end{figure}

 E.3-4 first used the social network of their department to observe the performance of and the difference across embedding models. This dataset comprises $324$ employees working in seven groups. As shown in Fig.~\ref{fig:cluster_transition}, the experts first examined the metric distribution in parallel coordinates in the cluster transition view and obtained an overview of the network, such as \textit{degree} distribution and \textit{within module degree} distribution. They then loaded the three embedding results from \textit{DeepWalk}, \textit{struc2vec}, and \textit{node2vec}. After sufficient iterations of \textit{t-SNE} projection, they observed that several clusters were formed in \textit{DeepWalk} and \textit{node2vec} space, while \textit{struc2vec} had a long and continuous ``tail''. The experts selected one cluster in \textit{DeepWalk} space and all the correspondences were highlighted and connected. \textit{DeepWalk} and \textit{node2vec} preserved the nodes in a cluster well, while \textit{struc2vec} dispersed these nodes. The experts were interested in knowing the position of the department leader; therefore, they clicked the corresponding node in the graph view (Fig.~\ref{fig:case0}(1. Click a node)). This node linked to several nodes in each cluster (indicated by arrows) other than its nearby neighbors. The experts then speculated that the long ``tail'' in \textit{struc2vec} projection space indicated the department hierarchy. Therefore, they lassoed a part of the ``tail'' and confirmed their hypothesis (Fig.~\ref{fig:case0}(2. Lasso nodes)). The interns were distributed at the end of the ``tail,'' and \textit{struc2vec} could accurately identify them. Another interesting finding was the different group statuses in the entire department (Fig.~\ref{fig:case0}(3. Lasso nodes)). ``\textit{What a pity! Our group is on the periphery of the department,}'' said E.3. This finding could be observed from the original graph, \textit{DeepWalk}, or \textit{node2vec} spaces. These observations confirmed the capability of \textit{struc2vec} to capture the structural identity~\cite{ribeiro2017struc2vec}, while \textit{DeepWalk} and \textit{node2vec} focused on the local neighboring structures.

\subsection{Case One: Verifying Preserved Node Metrics}
\noindent
This case uses the preceding \textit{csphd} dataset mentioned to verify the node metrics preserved by different embedding models.

\par
\begin{figure}[h]
    \centering
        \squeezeup
    \includegraphics[width=\linewidth]{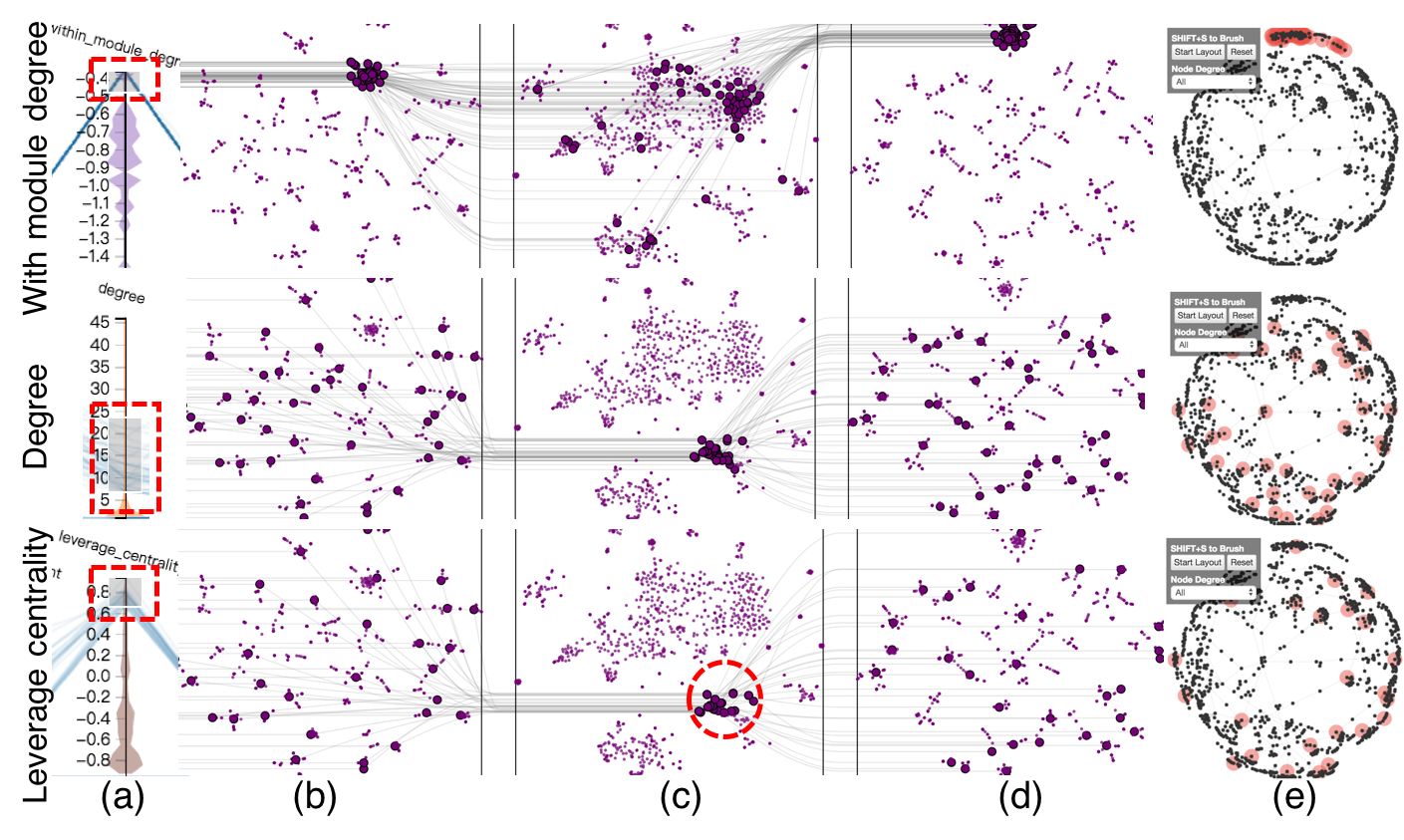} 
    \squeezeup
    \squeezeup
    \caption{Verify preserved metrics at the cluster level: (a) Filter nodes with similar \textit{within module degree}, \textit{degree}, \textit{leverage centrality} and observe distributions and transitions of the corresponding nodes in (b) \textit{DeepWalk}, (c) \textit{struc2vec}, (d) \textit{node2vec} (p=q=1) and (e) graph view.}
    \label{fig:case1}
    \squeezeup
\end{figure}

 \textbf{Verifying Preserved Metrics at Cluster Level.} The regression-based pairwise node metric analysis (Section 4.1) enables us to understand the correlation of metrics between a node-node pair in the graph space and the embedding space. For example, we have identified that the most preserved node metric by \textit{DeepWalk} is \textit{within module degree}, while \textit{degree} is the most preserved metric by \textit{struc2vec}, followed by \textit{leverage centrality}, etc. To verify these preserved metrics at the cluster level, as shown in Fig.~\ref{fig:case1}, we brushed several nodes from the axis of \textit{within module degree} and observed that the corresponding nodes were closely gathered in \textit{DeepWalk} and \textit{node2vec}; by contrast, \textit{struc2vec} split some of the nodes apart. Similarly, when we selected nodes with similar \textit{degree} or \textit{leverage centrality}, \textit{struc2vec} maintained the closeness of these nodes well, while the other spaces could not closely preserve nodes (R.1, R.3).

\par \textbf{Verifying Preserved Metrics at Instance Level.} We analyzed the neighboring information and compared the ranking performance of each embedding to further verify the preserved metrics at the instance level (R.4-5). We first lassoed a cluster of nodes and the corresponding nodes will be highlighted and linked across embedding spaces (Fig.~\ref{fig:teaser}(b1-e1)). We then select a ``hub node'' (label 20) in this cluster, and the top row in the pairwise ranking view shows its metric distribution (Fig.~\ref{fig:teaser}(a)). As shown in Fig.~\ref{fig:teaser}(d2), all the neighboring nodes had similar value distributions on \textit{leverage centrality}, \textit{participation coefficient}, \textit{knn}, \textit{PageRank}, and \textit{degree} in \textit{struc2vec}, while \textit{eccentricity} and \textit{within module degree} were preserved better by \textit{DeepWalk} (Fig.~\ref{fig:teaser}(c2)) and \textit{node2vec} (Fig.~\ref{fig:teaser}(e2)). We then highlighted the nodes of each ranking list in each \textit{t-SNE} projection space in the cluster transition view by clicking ``Filter'' button (Fig.~\ref{fig:teaser}(f)). Clearly, all the highlighted nodes in the lassoed cluster were captured by the neighboring ranking list of \textit{DeepWalk} (Fig.~\ref{fig:teaser}(c3)) and \textit{node2vec} (Fig.~\ref{fig:teaser}(e3)) because they all converged to a single cluster. However, most of them were not preserved closely by \textit{struc2vec} because they were split into several clusters (Fig.~\ref{fig:teaser}(d3)). \textit{DeepWalk} and \textit{node2vec} maintained the local neighborhoods of the selected node (Fig.~\ref{fig:teaser}(c3, e3)), while \textit{struc2vec} could maintain the ``hub'' nodes that are similar to the selected node together (Fig.~\ref{fig:teaser}(d3)), of which the Euclidean distance between them may be more volatile (Fig.~\ref{fig:teaser}(g)). The \textit{NDCG} scores of the rankings also confirmed our findings. We fixed the selected node (the top row of each ranking) and sorted all the other graph nodes using Euclidean distance with the selected node on the basis of the node metrics of \textit{leverage centrality}, \textit{participation coefficient}, \textit{PageRank}, \textit{degree}, and \textit{within module degree}. For example, when sorting by \textit{leverage centrality}, \textit{participation coefficient}, \textit{PageRank}, and \textit{degree} in the graph space, \textit{DeepWalk} and \textit{node2vec} return $0.109$, thereby indicating they fail to preserve well these metrics, compared with $0.6$-$0.8$ NDCG scores of \textit{struc2vec}. When sorting by \textit{within module degree}, \textit{struc2vec} fails to maintain the nodes with similar \textit{within module degree} close with NDCG score of $0.125$, while \textit{DeepWalk} and \textit{node2vec} return $0.556$.

\par \textbf{Summarizing the Takeaways.} The experts (E.1-2) obtained an intuitive understanding of the preserved node metrics by the showcased embedding models. For example, in the third row of Fig.~\ref{fig:case1}, E.1 observed that \textit{struc2vec} placed the nodes with high positive values of \textit{leverage centrality} together (indicated by the red circle): ``\textit{nodes with positive values of leverage centrality may influence their neighbors as the neighbors tend to have fewer connections with others.}'' They also found that \textit{participation coefficient} and \textit{PageRank} were well preserved by \textit{struc2vec}. \textit{Participation coefficient} and \textit{PageRank} are the metrics that specify whether nodes are truly ``hub'' nodes, while \textit{DeepWalk} and \textit{node2vec} preserve the nodes which are ``bound to local links'' reflected by \textit{within module degree}, ``\textit{these nodes do not play the role of connectors between clusters,}'' said E.2.

\subsection{Case Two: Understanding Hyper-Parameters}
\noindent
This case shows how \textit{EmbeddingVis} helped E.1-2 understand the effect of hyper-parameters $p$ and $q$ in \textit{node2vec} by using \textit{csphd} and a synthetic dataset. As claimed in~\cite{grover2016node2vec}, $p$ and $q$ control the searching strategy of random walk, i.e., a high value of $p$ will explore the network in a moderate manner, while a low value will lead the walk to backtrack a step and maintain the closeness of walk ``local'' to the starting node. Meanwhile, $q$ differentiates the searching between ``inward'' (\textit{BFS} with $q>1$) and ``outward'' (\textit{DFS} with $q<1$) nodes.

\begin{figure}[h]
    \centering
        \squeezeup
    \includegraphics[width=\linewidth]{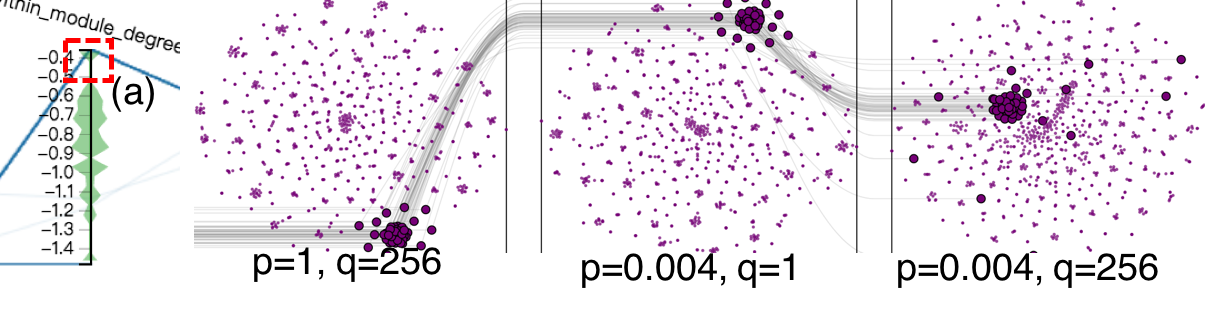} 
    \squeezeup
    \squeezeup
    \caption{Nodes with similar \textit{within module degree} are filtered and all identical ones are linked across three versions of \textit{node2vec} spaces.}
    \label{fig:case2_1}
    \squeezeup
\end{figure}

 \begin{figure*}[h]
    \centering
    \vspace{-5mm}
    \includegraphics[width=\linewidth]{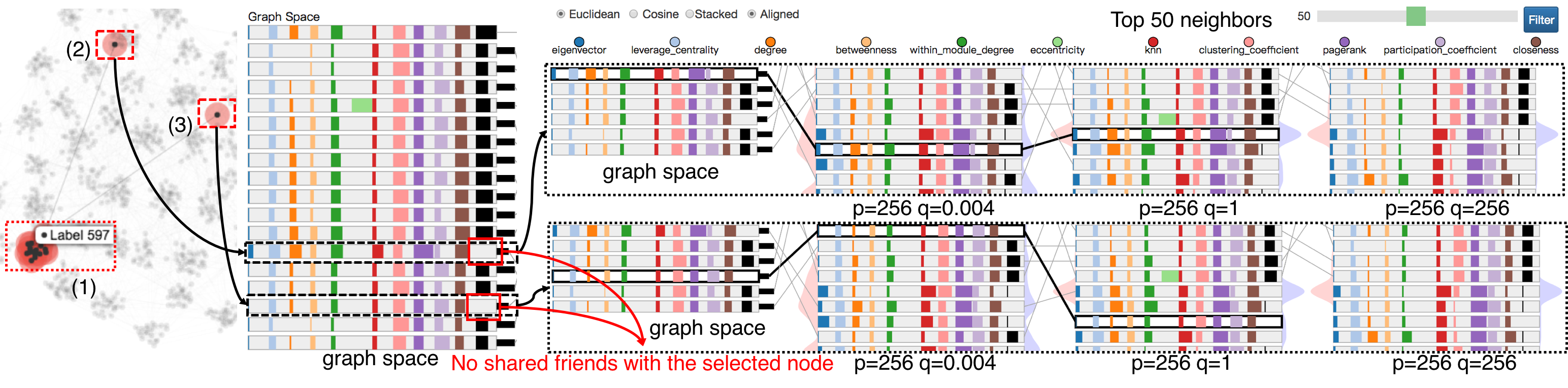} 
    \squeezeup
    \squeezeup
    \caption{(1) Click a node (label $597$) and generate its neighbor ranking lists for graph space and three \textit{node2vec} spaces. (2, 3) Two neighbors have no other links to the selected node's community. They are only captured by the first two versions in their top $50$ ranking lists.}
    \label{fig:case2_5}
    \squeezeup
\end{figure*}

\par \textbf{Revisiting Degrades Performance by $p$.} The regression analysis (Section 4.1) enables the experts to learn the average regression accuracy of the nine versions of \textit{node2vec}, among which six could achieve over 0.95 $R^2$ score. However, the versions of ($p=1$, $q=256$), ($p=0.004$, $q=1$) and ($p=0.004$, $q=256$) could achieve only approximately 0.7-0.8. Hence, the experts showed interests in these versions and loaded their corresponding embedding results. From the previous case study, E.2 obtained the insight that \textit{node2vec} preserved \textit{within module degree}; therefore, he filtered a small range of nodes with similar \textit{within module degree} (Fig.~\ref{fig:case2_1}(a)). The cluster transition view immediately responded to his operation and displayed the node transition across the three embedding spaces. He found that the embedding space with $p=0.004$ and $q=256$ had the worst performance; it did not maintain the nodes as closely as those in the two other spaces. However, compared with $p=q=1$ of \textit{node2vec} in Fig.~\ref{fig:case1}, all three versions failed to maintain the nodes sufficiently closely. The expert confirmed that ``\textit{revisiting an already visited node would limit the scope of the walks around the starting node, so the local neighboring nodes could not be preserved well.}''

\par \textbf{Differences Between \textit{DFS} and \textit{BFS} by $q$.} The experts shifted their attention to whether $q$ could significantly differentiate embedding results. They selected three groups of parameters, namely, ($p=256$, $q=0.004$), ($p=256$, $q=1$), and ($p=256$, $q=256$) for comparative analysis. However, after either inspecting \textit{t-SNE} embedding spaces or selecting nodes in graph view and comparing the pairwise neighbor rankings, they did not observe significant differences: ``\textit{I cannot explain this},'' said E.1. ``\textit{Can you find some structures, of which the connectivities inside a community are high but the community has some `bridge' nodes that link to other communities?}'' suggested by E.2. To narrow down the analysis, we generated a synthetic network utilizing the Barab{\'a}si-Albert model~\cite{albert2002statistical} and an algorithm proposed by Lancichinetti et al.~\cite{lancichinetti2008benchmark}, which specifically focuses on creating benchmark graphs with complex community structures and allows for generation of realistic synthetic networks. Fig.~\ref{fig:structure} and Fig.~\ref{fig:case2_5} show the analysis process for this dataset. We shifted to the structural view and clustered the network nodes into three (Fig.~\ref{fig:structure}(a)) on the basis of the seven structure-related metrics (Section 4.2) (R.2). We selected one cluster (Fig.~\ref{fig:structure}(b)) and observed its ``average distance vector'', which was generated by the three versions of \textit{node2vec}. At the beginning (Fig.~\ref{fig:structure}(c)), the distance of $p=256$ and $q=256$ version was lower than the distances from the other two versions; however, the distance exceeded them as the degree increased (Fig.~\ref{fig:structure}(d)). The experts were curious about this ``\textbf{turning point}.'' ``\textit{Interesting. No matter how I change the number of clusters, this always happens,}'' said E.2. The network degree varied from $12$ to $24$, thereby indicating that a node in this network had at least $12$ neighbors. We then turned to the graph view and clicked one node (Fig.~\ref{fig:case2_5}(1)), which was identified as having a tightly knit community comprising approximately $12$ neighbors and two distant nodes (Fig.~\ref{fig:case2_5}(2-3)) functioning as ``bridge nodes'' connecting to other communities. The two ``bridge nodes'' had no other links to the selected node community and shared no common friends with the selected node (indicated by red rectangles in Fig.~\ref{fig:case2_5}). When we hovered on them, we identified that the first two versions of \textit{node2vec} captured them but the third one failed to capture these nodes in its ranking list of top $50$ neighbors. We also attempted a few other nodes and discovered the same phenomenon. ``\textit{This version of node2vec conducts only a BFS-like (not BFS) strategy. Since the two `bridge nodes' have no links to any other nodes in the community, it has a great chance to miss them when $q$ is high,}'' said E.2.

\par \textbf{Summarizing the Takeaways.} The experts were convinced by what we identified through the analytical exploration. A small $p$ would easily lead to a frequent revisiting of an already visited node. However, the claimed difference that $q$ differentiates the random walk between \textit{BFS} and \textit{DFS} could not be easily observed. E.2 responded that ``\textit{the average distance vector in structural view helps us understand that a higher value of $q$ can keep the `focal node'-`neighborhood' structure much closer if they have a strong connectivity among neighbors of the focal node, but it tends to ignore those neighbors with no other links to the community}.'' ``\textit{In real-world networks, the structure is much more complicated and tuning $q$ alone could not significantly affect the embedding,}'' said E.1.

\section{Discussion}
\noindent
We conducted a half an hour semi-structured interview with our experts (E.1-4). We first asked them to evaluate the experimental results and the use of ``average distance vector'' to depict structural characteristics empirically. E.1-2 reported that ``\textit{the high accuracy of decision tree regression indicates that the embeddings preserve node metrics in a non-linear way.}'' They felt that ``\textit{the results make sense}'' after confirming the preserved metrics through interactive analysis. E.4 stated that ``\textit{it is practical to use embedding vectors to describe structural information since the trend of a cluster's `average distance vector' curve is similar across models.}''

\par \textbf{System Usability.} All the experts appreciated the capability of \textit{EmbeddingVis} to support interactive exploration and comparison among embedding models. E.1-2 suggested that our system would greatly boost their work efficiency. Conventionally, they have to manually sample some data they are familiar with, run the embedding algorithms on the example inputs, and then check how well the outputs match their intuition. ``\textit{Previously, we could only examine embeddings with our familiar networks. But now I can also explore other unfamiliar graphs.}'' E.2 described his past experience of comparing the neighbor rankings from different embedding models as trial-and-error. ``\textit{All I got each round was a single ranking score and could not dig deeper to find the underlying cause. This is quite stressful and time-consuming.}'' With our system, they could easily and intuitively compare interesting nodes/clusters and their distributions in different embedding spaces. ``\textit{I can explore different but interconnected information about a node/cluster in various views,}'' said E.2. They commented that ``\textit{EmbeddingVis is very useful because it provides a novel and highly interactive way to uncover the relationships between metrics and embedding vectors.}''

\par \textbf{Visual Design.} We drew inspiration from the observational study of the experts' conventional practices to inform the system design, such as analyzing the network pairwise/structural features, $2$D projection clustering analysis, and ranking measurement. We deliberately selected familiar visual metaphors so that the experts could quickly get accustomed to the visual encodings and designs. After being introduced to the basic views and functions of \textit{EmbeddingVis}, they developed a path through the system for exploration.

\par \textbf{Generalizability.} We discussed with the experts (E.3-4) which component(s) of our system can be directly applied to other analytical scenarios and which one(s) need customization to further explore the potential of \textit{EmbeddingVis}. Two findings were obtained: (1) \textit{Visual Design Generalization.} E.3 mentioned that our system is already very general in terms of the metrics incorporated. But if we allow users to modify the metrics list, the system would be very applicable to other scenarios. ``\textit{I would personally like to add a few business-related metrics derived from historical data to the system, e.g., whether this user has ever clicked on a targeted advertisement,}'' said E.3. (2) \textit{Embedding Model Generalization.} Although we only showcase three embedding models in this paper, other types, e.g., \textit{LINE}~\cite{tang2015line}, \textit{SDNE}~\cite{wang2016structural} can be included with simple configuration.

\par \textbf{Scalability.} First, we have used more than ten colors to differentiate metrics and keep the color encoding consistent across all the views. However, we are fully aware that only a small number of colors can be effectively used as category labels~\cite{ware2012information}. If there are more metrics, then we should support dynamic filtering to display the desired metrics. Second, in this paper, considering the time consumed in generating the metrics and the embedding results, we only use networks with $1000$ to $3000$ nodes for demonstration. For a larger network, we can sample the nodes or use \textit{GPU} to accelerate the computing process. We also envision the scalability issues with the visualization when dealing with larger network data. Two findings arose: \textit{(1) Remove iteration process.} Currently, we leverage \textit{t-SNE} and directly visualize its iterative process in the front-end. When dealing with larger networks, we should remove the iterative process to the backend and only preserve the coordinates of points in the final step. \textit{(2) Replace with canvas.} Switching the current SVG-based visualization to canvas-based \textit{WebGL} rendering could significantly boost the visualization performance for larger networks.

\par \textbf{Limitation.} We may fail to capture a few metrics with a possibly high correlation between the graph space and the embedding space due to limited datasets we used or to weigh the feature importance of node metrics by averaging over all datasets. We focused only on two hyper-parameters with limited settings and maintained the similarity of other parameters, such as \textit{window size}, and we only worked with a small team of experts (E.1-4) in the evaluation and thus could not provide a quantitative assessment of the system.

\section{Conclusion and Future Work}
\noindent
In this paper, we propose \textit{EmbeddingVis} to enable a comparative inspection of embedding vectors at the cluster, instance, and structural levels. Case studies and experts' feedback verify the efficacy of our system. Several ways are available to improve the system. First, we plan to perform a systematic evaluation that involves additional experts. We can then iteratively refine the system based on the feedback. Second, we will involve more embedding models and more specific tasks. Third, we also want to study how a model changes over time, i.e., how the performance of an embedding model changes over the iterative process and when it reaches a sufficiently good result. Currently, we only work with the final embedding results after a fixed number of iterations.

\acknowledgments{
We are grateful for the valuable feedback and comments provided by Prof. Huamin Qu and the anonymous reviewers. This research was supported in part by WeChat-HKUST Joint Lab on AI Technology (WHAT LAB) grant\#1617170-0 and HK RGC GRF 16213317.}

\bibliographystyle{abbrv-doi}

\bibliography{template}

\begin{thebibliography}{10}

\bibitem{albert2002statistical}
R.~Albert and A.-L. Barab{\'a}si.
\newblock Statistical mechanics of complex networks.
\newblock {\em Reviews of modern physics}, 74(1):47, 2002.

\bibitem{alper2013weighted}
B.~Alper, B.~Bach, N.~Henry~Riche, T.~Isenberg, and J.-D. Fekete.
\newblock Weighted graph comparison techniques for brain connectivity analysis.
\newblock In {\em Proceedings of the SIGCHI conference on human factors in
  computing systems}, pp. 483--492. ACM, 2013.

\bibitem{armstrong2011illusions}
J.~Armstrong.
\newblock Illusions in regression analysis.
\newblock 2011.

\bibitem{berlingerio2012netsimile}
M.~Berlingerio, D.~Koutra, T.~Eliassi-Rad, and C.~Faloutsos.
\newblock Netsimile: A scalable approach to size-independent network
  similarity.
\newblock {\em arXiv preprint arXiv:1209.2684}, 2012.

\bibitem{cai2017comprehensive}
H.~Cai, V.~W. Zheng, and K.~C.-C. Chang.
\newblock A comprehensive survey of graph embedding: Problems, techniques and
  applications.
\newblock {\em arXiv preprint arXiv:1709.07604}, 2017.

\bibitem{cao2015grarep}
S.~Cao, W.~Lu, and Q.~Xu.
\newblock Grarep: Learning graph representations with global structural
  information.
\newblock In {\em Proceedings of the 24th ACM International on Conference on
  Information and Knowledge Management}, pp. 891--900. ACM, 2015.

\bibitem{chang2015heterogeneous}
S.~Chang, W.~Han, J.~Tang, G.-J. Qi, C.~C. Aggarwal, and T.~S. Huang.
\newblock Heterogeneous network embedding via deep architectures.
\newblock In {\em Proceedings of the 21th ACM SIGKDD International Conference
  on Knowledge Discovery and Data Mining}, pp. 119--128. ACM, 2015.

\bibitem{collins2007vislink}
C.~Collins and S.~Carpendale.
\newblock Vislink: Revealing relationships amongst visualizations.
\newblock {\em IEEE Transactions on Visualization and Computer Graphics},
  13(6):1192--1199, 2007.

\bibitem{gleicher2011visual}
M.~Gleicher, D.~Albers, R.~Walker, I.~Jusufi, C.~D. Hansen, and J.~C. Roberts.
\newblock Visual comparison for information visualization.
\newblock {\em Information Visualization}, 10(4):289--309, 2011.

\bibitem{gratzl2013lineup}
S.~Gratzl, A.~Lex, N.~Gehlenborg, H.~Pfister, and M.~Streit.
\newblock Lineup: Visual analysis of multi-attribute rankings.
\newblock {\em IEEE transactions on visualization and computer graphics},
  19(12):2277--2286, 2013.

\bibitem{grover2016node2vec}
A.~Grover and J.~Leskovec.
\newblock node2vec: Scalable feature learning for networks.
\newblock In {\em Proceedings of the 22nd ACM SIGKDD international conference
  on Knowledge discovery and data mining}, pp. 855--864. ACM, 2016.

\bibitem{gu2017hidden}
W.~Gu, L.~Gong, X.~Lou, and J.~Zhang.
\newblock The hidden flow structure and metric space of network embedding
  algorithms based on random walks.
\newblock {\em Scientific reports}, 7(1):13114, 2017.

\bibitem{heimerl2018embeddings}
F.~Heimerl and M.~Gleicher.
\newblock Interactive analysis of word vector embeddings.
\newblock In {\em Computer Graphics Forum}, vol.~37. Wiley Online Library,
  2018.

\bibitem{hsu2008acceptance}
C.-L. Hsu and J.~C.-C. Lin.
\newblock Acceptance of blog usage: The roles of technology acceptance, social
  influence and knowledge sharing motivation.
\newblock {\em Information \& management}, 45(1):65--74, 2008.

\bibitem{huang2017label}
X.~Huang, J.~Li, and X.~Hu.
\newblock Label informed attributed network embedding.
\newblock In {\em Proceedings of the Tenth ACM International Conference on Web
  Search and Data Mining}, pp. 731--739. ACM, 2017.

\bibitem{inselberg1987parallel}
A.~Inselberg and B.~Dimsdale.
\newblock Parallel coordinates for visualizing multi-dimensional geometry.
\newblock In {\em Computer Graphics 1987}, pp. 25--44. Springer, 1987.

\bibitem{kim2016pixelsne}
M.~Kim, M.~Choi, S.~Lee, J.~Tang, H.~Park, and J.~Choo.
\newblock Pixelsne: Visualizing fast with just enough precision via
  pixel-aligned stochastic neighbor embedding.
\newblock {\em arXiv preprint arXiv:1611.02568}, 2016.

\bibitem{labrini2015graph}
H.~Labrini.
\newblock Graph-based model for distribution systems: Application to planning
  problem.
\newblock Master's thesis, University of Waterloo, 2015.

\bibitem{lancichinetti2008benchmark}
A.~Lancichinetti, S.~Fortunato, and F.~Radicchi.
\newblock Benchmark graphs for testing community detection algorithms.
\newblock {\em Physical review E}, 78(4):046110, 2008.

\bibitem{lawrence1999digital}
S.~Lawrence, C.~L. Giles, and K.~Bollacker.
\newblock Digital libraries and autonomous citation indexing.
\newblock {\em Computer}, 32(6):67--71, 1999.

\bibitem{le2014probabilistic}
T.~M. Le and H.~W. Lauw.
\newblock Probabilistic latent document network embedding.
\newblock In {\em Proceedings of IEEE International Conference on Data Mining
  (ICDM)}, pp. 270--279. IEEE, 2014.

\bibitem{li2017attributed}
J.~Li, H.~Dani, X.~Hu, J.~Tang, Y.~Chang, and H.~Liu.
\newblock Attributed network embedding for learning in a dynamic environment.
\newblock In {\em Proceedings of the 2017 ACM on Conference on Information and
  Knowledge Management}, pp. 387--396. ACM, 2017.

\bibitem{li2017visual}
Q.~Li, Q.~Shen, Y.~Ming, P.~Xu, Y.~Wang, X.~Ma, and H.~Qu.
\newblock A visual analytics approach for understanding egocentric intimacy
  network evolution and impact propagation in mmorpgs.
\newblock In {\em Proceedings of IEEE Pacific Visualization Symposium
  (PacificVis)}, pp. 31--40. IEEE, 2017.

\bibitem{li2018multi}
Q.~Li, Z.~Wu, P.~Xu, H.~Qu, and X.~Ma.
\newblock A multi-phased co-design of an interactive analytics system for moba
  game occurrences.
\newblock In {\em Proceedings of the 2018 on Designing Interactive Systems
  Conference 2018}, pp. 1321--1332. ACM, 2018.

\bibitem{livisual2017}
Q.~Li, P.~Xu, Y.~Y. Chan, Y.~Wang, Z.~Wang, H.~Qu, and X.~Ma.
\newblock A visual analytics approach for understanding reasons behind
  snowballing and comeback in moba games.
\newblock {\em IEEE transactions on visualization and computer graphics},
  23(1):211--220, 2017.

\bibitem{liao2017attributed}
L.~Liao, X.~He, H.~Zhang, and T.-S. Chua.
\newblock Attributed social network embedding.
\newblock {\em arXiv preprint arXiv:1705.04969}, 2017.

\bibitem{lin2015learning}
Y.~Lin, Z.~Liu, M.~Sun, Y.~Liu, and X.~Zhu.
\newblock Learning entity and relation embeddings for knowledge graph
  completion.
\newblock In {\em AAAI}, vol.~15, pp. 2181--2187, 2015.

\bibitem{liu2018deeptracker}
D.~Liu, W.~Cui, K.~Jin, Y.~Guo, and H.~Qu.
\newblock Deeptracker: Visualizing the training process of convolutional neural
  networks.
\newblock {\em To appear in ACM Transactions on Intelligent Systems and
  Technology}, 2018.

\bibitem{liu2018visual}
S.~Liu, P.-T. Bremer, J.~J. Thiagarajan, V.~Srikumar, B.~Wang, Y.~Livnat, and
  V.~Pascucci.
\newblock Visual exploration of semantic relationships in neural word
  embeddings.
\newblock {\em IEEE transactions on visualization and computer graphics},
  24(1):553--562, 2018.

\bibitem{maaten2008visualizing}
L.~v.~d. Maaten and G.~Hinton.
\newblock Visualizing data using t-sne.
\newblock {\em Journal of machine learning research}, 9(Nov):2579--2605, 2008.

\bibitem{mclachlan2008liverac}
P.~McLachlan, T.~Munzner, E.~Koutsofios, and S.~North.
\newblock Liverac: interactive visual exploration of system management
  time-series data.
\newblock In {\em Proceedings of the SIGCHI Conference on Human Factors in
  Computing Systems}, pp. 1483--1492. ACM, 2008.

\bibitem{mikolov2013efficient}
T.~Mikolov, K.~Chen, G.~Corrado, and J.~Dean.
\newblock Efficient estimation of word representations in vector space.
\newblock {\em arXiv preprint arXiv:1301.3781}, 2013.

\bibitem{nishana2013graph}
S.~Nishana and S.~Surendran.
\newblock Graph embedding and dimensionality reduction-a survey.
\newblock {\em International Journal of Computer Science \& Engineering
  Technology (IJCSET)}, 4(1):29--34, 2013.

\bibitem{ou2016asymmetric}
M.~Ou, P.~Cui, J.~Pei, Z.~Zhang, and W.~Zhu.
\newblock Asymmetric transitivity preserving graph embedding.
\newblock In {\em Proceedings of the 22nd ACM SIGKDD international conference
  on Knowledge discovery and data mining}, pp. 1105--1114. ACM, 2016.

\bibitem{parberry1995sigact}
I.~Parberry and D.~S. Johnson.
\newblock The sigact theoretical computer science genealogy: Preliminary
  report, 1995.

\bibitem{perozzi2014deepwalk}
B.~Perozzi, R.~Al-Rfou, and S.~Skiena.
\newblock Deepwalk: Online learning of social representations.
\newblock In {\em Proceedings of the 20th ACM SIGKDD international conference
  on Knowledge discovery and data mining}, pp. 701--710. ACM, 2014.

\bibitem{rauber2017visualizing}
P.~E. Rauber, S.~G. Fadel, A.~X. Falcao, and A.~C. Telea.
\newblock Visualizing the hidden activity of artificial neural networks.
\newblock {\em IEEE transactions on visualization and computer graphics},
  23(1):101--110, 2017.

\bibitem{ribeiro2017struc2vec}
L.~F. Ribeiro, P.~H. Saverese, and D.~R. Figueiredo.
\newblock struc2vec: Learning node representations from structural identity.
\newblock In {\em Proceedings of the 23rd ACM SIGKDD International Conference
  on Knowledge Discovery and Data Mining}, pp. 385--394. ACM, 2017.

\bibitem{salehi2017properties}
F.~Salehi~Rizi, M.~Granitzer, and K.~Ziegler.
\newblock Properties of vector embeddings in social networks.
\newblock {\em Algorithms}, 10(4):109, 2017.

\bibitem{sen2008collective}
P.~Sen, G.~Namata, M.~Bilgic, L.~Getoor, B.~Galligher, and T.~Eliassi-Rad.
\newblock Collective classification in network data.
\newblock {\em AI magazine}, 29(3):93, 2008.

\bibitem{shneiderman2003eyes}
B.~Shneiderman.
\newblock The eyes have it: A task by data type taxonomy for information
  visualizations.
\newblock In {\em The Craft of Information Visualization}, pp. 364--371.
  Elsevier, 2003.

\bibitem{smilkov2016embedding}
D.~Smilkov, N.~Thorat, C.~Nicholson, E.~Reif, F.~B. Vi{\'e}gas, and
  M.~Wattenberg.
\newblock Embedding projector: Interactive visualization and interpretation of
  embeddings.
\newblock {\em arXiv preprint arXiv:1611.05469}, 2016.

\bibitem{tang2015line}
J.~Tang, M.~Qu, M.~Wang, M.~Zhang, J.~Yan, and Q.~Mei.
\newblock Line: Large-scale information network embedding.
\newblock In {\em Proceedings of the 24th International Conference on World
  Wide Web}, pp. 1067--1077. International World Wide Web Conferences Steering
  Committee, 2015.

\bibitem{tian2014learning}
F.~Tian, B.~Gao, Q.~Cui, E.~Chen, and T.-Y. Liu.
\newblock Learning deep representations for graph clustering.
\newblock In {\em AAAI}, pp. 1293--1299, 2014.

\bibitem{ugander2012structural}
J.~Ugander, L.~Backstrom, C.~Marlow, and J.~Kleinberg.
\newblock Structural diversity in social contagion.
\newblock {\em Proceedings of the National Academy of Sciences},
  109(16):5962--5966, 2012.

\bibitem{wang2016structural}
D.~Wang, P.~Cui, and W.~Zhu.
\newblock Structural deep network embedding.
\newblock In {\em Proceedings of the 22nd ACM SIGKDD international conference
  on Knowledge discovery and data mining}, pp. 1225--1234. ACM, 2016.

\bibitem{wang2013theoretical}
Y.~Wang, L.~Wang, Y.~Li, D.~He, W.~Chen, and T.-Y. Liu.
\newblock A theoretical analysis of ndcg ranking measures.
\newblock In {\em Proceedings of the 26th Annual Conference on Learning Theory
  (COLT 2013)}, 2013.

\bibitem{ware2012information}
C.~Ware.
\newblock {\em Information visualization: perception for design}.
\newblock Elsevier, 2012.

\bibitem{wu2016egoslider}
Y.~Wu, N.~Pitipornvivat, J.~Zhao, S.~Yang, G.~Huang, and H.~Qu.
\newblock egoslider: Visual analysis of egocentric network evolution.
\newblock {\em IEEE transactions on visualization and computer graphics},
  22(1):260--269, 2016.

\bibitem{xiao2017ssp}
H.~Xiao, M.~Huang, L.~Meng, and X.~Zhu.
\newblock Ssp: Semantic space projection for knowledge graph embedding with
  text descriptions.
\newblock In {\em AAAI}, vol.~17, pp. 3104--3110, 2017.

\bibitem{yanardag2015deep}
P.~Yanardag and S.~Vishwanathan.
\newblock Deep graph kernels.
\newblock In {\em Proceedings of the 21th ACM SIGKDD International Conference
  on Knowledge Discovery and Data Mining}, pp. 1365--1374. ACM, 2015.

\bibitem{yin2017local}
H.~Yin, A.~R. Benson, J.~Leskovec, and D.~F. Gleich.
\newblock Local higher-order graph clustering.
\newblock In {\em Proceedings of the 23rd ACM SIGKDD International Conference
  on Knowledge Discovery and Data Mining}, pp. 555--564. ACM, 2017.

\end{thebibliography}
\end{document}